\documentclass[prb,aps,twocolumn,groupedaddress,12pts]{revtex4-1}
\usepackage{epsfig,latexsym,amssymb,amsmath,amsbsy,graphics,graphicx,bm,natbib}

\def \beq{\begin{equation}}
\def \eeq{\end{equation}}
\def \beqarr{\begin{eqnarray}}
\def \eeqarr{\end{eqnarray}}

\begin{document}

\title{Quantum Hall Effects in Graphene-Based Two-Dimensional Electron Systems}

\author{Yafis Barlas}
\author{Kun Yang}
\affiliation{National High Magnetic Field Laboratory and Department of Physics, Florida State
University, FL 32306, USA}

\author{A. H. MacDonald}
\affiliation{Department of Physics, University of Texas at Austin,
Austin, TX, 78712, USA}

\date{\today}

\begin{abstract}

In this article we review the quantum Hall physics of graphene based two-dimensional electron systems,
with a special focus on recent experimental and theoretical developments.
We explain why graphene and bilayer graphene can be viewed
respectively as $J=1$ and $J=2$ chiral two-dimensional electron gases (C2DEGs), and why this property
frames their quantum Hall physics.
The current status of experimental and theoretical work on the role of electron-electron interactions
is reviewed at length with an emphasis on unresolved issues in the field, including assessing the role
of disorder in current experimental results.
Special attention is given to the interesting low magnetic field limit and to the relationship between
quantum Hall effects and the spontaneous anomalous Hall effects that might
occur in bilayer graphene systems in the absence of a magnetic field.
\end{abstract}

\maketitle

\section{Introduction}

The quantum Hall effect (QHE) is one of the most remarkable phenomena in condensed matter
physics.~\cite{iqheexp,iqheth,fqheexp,fqheth}  From a phenomenological point of view it is a transport anomaly that is reinforced by
disorder-induced localization of charged quasiparticles.  It
is observed in two-dimensional electron systems whenever a charge gap, {\em i.e.} a jump in the chemical potential,
occurs at a density that is magnetic field dependent.
The gap-density's magnetic field dependence can be directly related to
Hall transport properties,\cite{halperinedge,streda} and to topological Chern indices that
characterize the dependence of single electron wave functions on Bloch wavevectors\cite{addreftknn} in
the special case of periodic systems, and many electron wave functions
on boundary condition angles in the general case.\cite{addrefniu,sheng03}
The Chern indices determine the values of the
Hall conductivity~\cite{sds,allanqhreview} which are quantized in units of the
conductance quantum $e^2/h$ in non-interacting electron systems with
fractional values allowed in interacting systems.

The simplest quantum Hall gaps follow directly from the single-particle physics of
kinetic energy quantization and Landau level formation in
two-dimensional systems, but interaction induced gaps are also common.
Landau quantization gaps can occur only at integer values of the
Landau level filling factor $\nu$.   Interaction induced gaps can occur at integer or
fractional values of $\nu$.  In a system with no periodic potential, and therefore
no length scale\cite{atomiccaveat} other than that defined by the magnetic flux quantum, gaps always
occur at fixed $\nu$ and have Hall conductance $\nu e^2/h$.
Some of the incompressible states associated with interaction induced gaps have exotic properties,\cite{fqheexp,fqheth,sds} including
quasiparticles with fractional charge and fractional, or even non-Abelian quantum statistics.
The discovery of the quantum Hall effect in the 1980's was a watershed event in modern physics.
The observation of perfectly quantized transport coefficients in a disordered system containing millions of electrons
was not only a great example of nature's tendency to surprise,
but also pointed the community to\cite{addrefgirvinahm} a new type of
order - topological order - which can occur in condensed matter.

Quantum Hall effects have been studied for several decades in the
two dimensional electron gases (2DEGs) found at interfaces between
different semiconductors and at interfaces between a semiconductor and an insulator.
These systems continue to be actively investigated~\cite{nonabelianrmp} and to reveal
surprises.  Semiconductor heterojunction 2DEGs are not
two-dimensional in the atomic sense since their electronic charge is typically spread
over $30$ or more atomic layers, but in the electronic sense because
their transverse coordinate degree of freedom is frozen out by
finite-size quantization gaps. The recent isolation of graphene,~\cite{phystodaygraphene,rmpgraphene} 
a two-dimensional honeycomb lattice of carbon atoms,
has enabled exploration~\cite{graphene_qhe1,graphene_qhe2,graphene_qhfexp} of the quantum Hall effect, 
along with many other electronic properties,
in a 2DEG that has a thickness of one atomic layer and is qualitatively distinct.
Because one, two, and three layer graphene systems stacked in different ways
have quite distinct electronic properties, graphene research
has given rise to an entire family of 2DEG systems with a very rich set of
behaviors.  This review will focus for the most part
on new questions that have arisen in experimental and theoretical
studies of the quantum Hall effect in one and two layer graphene systems.

The article is organized as follows.  In section II we briefly summarize some of the main
properties of graphene-based 2DEGs.
In section III we discuss the low-energy effective theories of graphene and bilayer graphene starting from
the nearest neighbor tight-binding model.   In the process we point out that the low-energy effective theories
of single layer and bilayer graphene are particular cases of a broader class which we refer to
as chiral 2DEG models.  Other members of this class of Hamiltonians are approximately realized in
thicker few layer graphene systems. In section IV we analyze the properties of chiral 2DEGs in the presence of a magnetic field, with special focus on their unusual Landau quantization pattern.
In section V we focus on interaction driven quantum Hall effects in graphene and bilayer graphene at
integer filling factors, focusing on the unusual $ \nu =0 $ QH plateau of both monolayer and bilayer graphene.
In section VI we address the recently discovered fractional quantum Hall (FQH) states in graphene and
offer some speculations on the physics that determines which fractional states are strongest.
Finally we conclude in section VII with some speculations on interesting directions for this rapidly evolving field.

There are already a number of excellent reviews~\cite{rmpgraphene,kunsreview,sdsreview,goerbigreview} on different aspects of the QHE in graphene, and there is bound to be some overlap of this
review with others in the literature.  The focus of this contribution is on the the Landau quantization pattern of
graphene and few layer graphene chiral 2DEG systems, and on the rich family of interaction dominated quantum Hall states
to which it leads.

\section{Graphene Quantum Hall Effect Primer}

\begin{figure*}[t]
\begin{center}
\includegraphics[width=7in,height=2.2in]{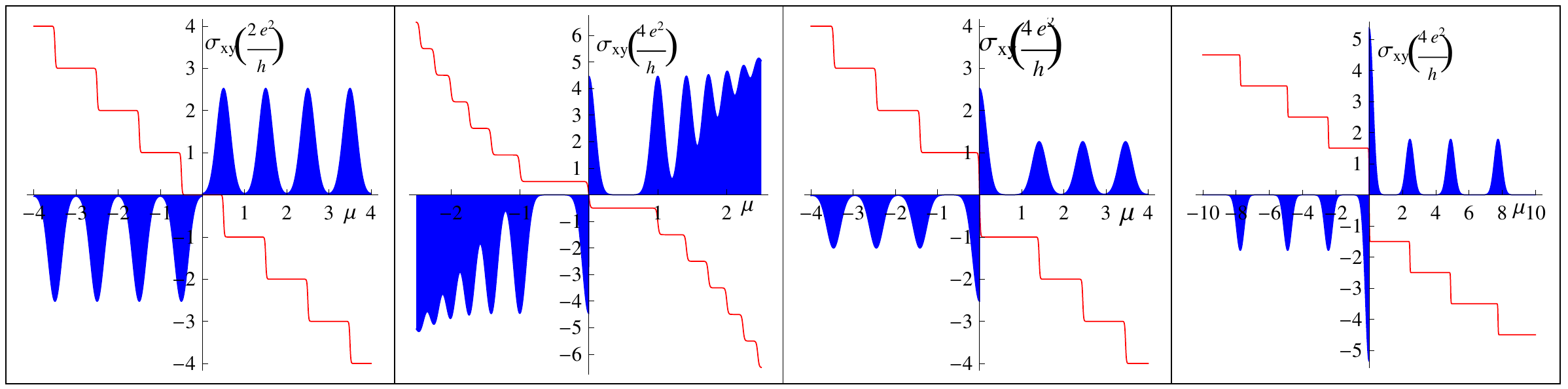}
\caption{Schematic illustration of the integer QH effect found in (a) conventional 2DEGs, (b) monolayer graphene, (c) bilayer graphene and (d) ABC stacked trilayer graphene as a function of chemical potential $\mu$ measured from the carrier neutrality point for fixed magnetic field.
The Hall conductivity $\sigma_{xy}$ is shown in red (solid) line and the longitudinal conductivity in blue.  Plateaus in these C2DEG systems appear at $4(n+J/2)e^2/h$ integer
multiples of $e^2/h$, where $n$ is an integer and $J$ determines the 2DEG chirality (see text for details). For monolayer, bilayer, and ABC stacked trilayer graphene $J$ is equal to the number of layers. The chemical potential $\mu$ is measured in units of $\hbar^2/(ml_{B}^2)$ for conventional 2DEGs and in units of $1/\gamma_{1}^{J-1}(\sqrt{2}\hbar v_{D}/l_{B})^J$ for a $J$-chiral 2DEG (see text for details). Note that Landau level separation increases with energy for $J=3$, decreases with energy for $J=1$ and is independent of energy for $J=2$. The shaded blue region corresponds to the disorder broadened Landau level density of states, which is inverted for holes carriers to facilitate clarity. The disorder broadened Landau level density of states is roughly the same form as the longitudinal conductivity except that its peaks are typically less sharp.  Because graphene is a gapless atomically thin 2DEG system its carrier density can be varied over a large range surrounding the carrier neutrality point.}
\label{figone}
\end{center}
\end{figure*}

Graphene can be viewed as a single layer of carbon atoms that has been isolated from a bulk graphite crystal.
Its electronic structure\cite{rmpgraphene} consists of $sp^{2}$ bonding and anti-bonding
bands which are located far from the Fermi energy and irrelevant for transport, and two
$\pi$ bands - one for each of the two $\pi$-orbitals per unit cell in the
material's bipartite honeycomb lattice.
In a neutral graphene sheet the lower energy $\pi$ valence band is full and the higher energy $\pi$ conduction band is empty.  The gap between conduction and valence bands vanishes at two points in the
Brillouin-zone, the inequivalent honeycomb lattice Brillouin-zone corners.
Only states located at nearby momenta are relevant for transport.
As described more fully in the next section,
the quantum mechanics of these states
can be accurately described over a remarkably wide energy window
using a ${\bf k} \cdot {\bf p}$ envelope function
Hamiltonian which has the form of a two-dimensional massless Dirac equation
in which the role of spin is played by the two-valued {\em which sublattice} degree of freedom.  (Because there are
two inequivalent momenta at which the gap vanishes, electrons also carry a two-valued
{\em valley} degree-of-freedom when described using the continuum ${\bf k} \cdot {\bf p}$
theory in addition to their pseudospin and real spin degrees-of-freedom.)  A number of interesting
properties of $\pi$-electrons in graphene follow from this ultra-relativistic quantum
mechanics, including the absence of a gap, linear dispersion at low energies, and momentum
space Berry phases. \\
\indent
In graphene, Landau level formation leads to unusual Hall plateaus on which the
Hall conductivity is quantized~\cite{gusynin} at half-odd-integer multiples (per valley and spin!)
of $e^2/h$.  The early~\cite{graphene_qhe1,graphene_qhe2} observation of this property in graphene 2DEGs
was an important demonstration that the experimental systems behaved nearly ideally
as expected.  When spin and valley degrees of freedom are included and
Zeeman coupling is neglected,Landau level formation in graphene leads
to quantized Hall conductivities with values
$\sigma_{xy} = \nu e^2/h$ with $\nu = \pm 4(n+1/2) = \pm 2, \pm 6, \pm 10, \cdots$.
The strongest quantum Hall effects in graphene occur at these filling factors.
Linear band dispersion in the absence of a magnetic field leads to Landau levels with
energies (measured from the band-crossing Dirac point) that are proportional to $\pm \sqrt{N}$ (see Fig. \ref{figone}).
For this reason the Landau levels close to the Dirac point
are more widely spaced allowing the quantum Hall effect at $\nu=\pm 2$ to persist to room temperatures
at very strong magnetic fields.~\cite{roomtemphall}  Because of the availability of graphene 2DEG systems,
the quantum Hall effect is no longer solely a low-temperature physics topic!
Similarly the quantum Hall effect at $\nu=\pm 2$ has been observed at low temperatures
at magnetic fields strengths well below 1 Tesla.

As sample qualities improved experiments began to reveal a large number of additional
Hall plateaus~\cite{graphene_qhfexp} {\em not} expected from Landau quantization alone,
the plateaus associated with these gaps could only be induced by electron-electron interactions.
Historically these gaps had been theoretically anticipated from two different directions.
The take off point for one group of researchers~\cite{nomura,gmd,aliceafisher,ydm,abanin}
had been the tendency, already familiar from semiconductor 2DEG work,
toward gaps at intermediate integer filling factors whenever a group of
Landau levels are degenerate either accidentally or because of a Hamiltonian symmetry.
The tendency for interactions in this case to generate mass terms in quasiparticle Hamiltonians
that open gaps, breaking symmetries if necessary, can be understood in terms of the
degeneracy of the Landau levels and the role of exchange interactions.\cite{jungwirthqhf}
A common example of this type of interaction induced gap is spontaneous spin-polarization~\cite{sondhi,fogler}
that is accompanied by a large exchange gap between Landau levels.  (The Zeeman splitting is often very
weak compared to the typical electron-electron interaction scale so that these gaps are essentially
spontaneous.)  For this reason the incompressible states associated with gaps that lift
Landau level degeneracies are often referred to as quantum Hall ferromagnets, even when
the gap in question is not a spin gap.  Another group of researchers\cite{gmss1,herbut} anticipated gaps at
filling factor $\nu=0$ in graphene sheets based on the idea, referred to as {\em magnetic catalysis}, that
these spontaneous mass gaps nearly occur in the absence of a magnetic field and are then catalyzed by
Dirac-point Landau level degeneracies.  Recent experiments seem to favor the
quantum Hall ferromagnetism scenario~\cite{graphene_fqhe1,graphene_fqhe2,BNsubstrate} since
interaction induced gaps have been observed at many integer filling factors, not only at $\nu=0$, in both single-layer and
bilayer graphene.  Interaction induced gaps were also anticipated at fractional
Landau level filling factors, the fractional quantum Hall effect (FQHE),
and have recently been revealed~\cite{graphene_fqhe1,graphene_fqhe2}
by transport measurements in suspended graphene samples and in graphene on hexagonal Boron Nitride~\cite{BNsubstrate} (h-BN) substrates.\\
\indent
Like single-layer graphene (see Fig.\ref{figtwo} for lattice structure), Bernal stacked bilayer graphene is a gapless semiconductor.
(The most common structure of bulk graphite consists of stacked Bernal bilayers.) As shown in Fig. \ref{figthree}
Bernal bilayers are characterized by a relative displacement between adjacent honeycomb lattices,
which places one of the sublattices in the top layer directly above the centers of the honeycombs of the layer below, whereas the other sublattice lies directly above a lower layer sublattice. Interlayer hopping between the closely spaced sites drives the energies of orbitals concentrated on the latter sites away from the Dirac point. The low energy degrees of freedom are associated with the remaining sites, the ones without a near-neighbor in the other layer. The low energy effective theory of bilayer graphene consequently also has a two-valued
low-energy {\em which sublattice} degree-of-freedom, one which applies however only below the
interlayer hopping energy scale $\sim 0.3 {\rm eV}$.   (The model of a bilayer as consisting of two coupled two-component 2D Dirac systems applies up to a much higher energy scale
$\sim 3 {\rm eV}$.) Since the two low-energy sublattices are localized in different layers, in the bilayer case the model's pseudospin can also be viewed equally well as a {\em which layer} degree-of-freedom that is readily coupled to an experimental probe simply by applying an electric field perpendicular to the
bilayer surface.\\
\indent
The low energy conduction and valence bands in bilayer graphene have parabolic dispersion~\cite{mccannfalko}
and a Berry phase of $2\pi$, twice the value of Dirac fermions. In a magnetic field, bilayer graphene's Landau level quantization
leads to another distinct integer quantum Hall effect (IQHE)~\cite{bg_iqhe,mccannfalko} with quantized
Hall conductivities  $\sigma_{xy} = \nu e^2/h $ at $\nu = \pm 4 (n+1) = \pm 4, \pm8, \pm 12 \cdots $.
The  eightfold degeneracy of the zero-energy $N=0$ Landau level is due to the inclusion of both $n=0 $ and $n=1$
orbital states among the zero energy solutions of the Dirac equation.
This additional degeneracy, along with valley and approximate spin degeneracy, leads to a set of
eight nearly degenerate Landau levels which are ripe for quantum Hall ferromagnetism, the $N=0$ octet.
Although investigations of electron-electron interaction effects in bilayer graphene
are still in their infancy, recent experiments in high magnetic field have clearly demonstrated
additional plateaus~\cite{bg_qhf1,bg_qhf2} due to electron-electron interactions.~\cite{ourbilayerqhf}

Because its $N=0$ states in different valleys are localized in different layers,
bilayer graphene is partially analogous to double layer quantum Hall systems
in semiconductors with two quantum wells.  These systems have exhibited
unusual transport anomalies due to excitonic superfluidity~\cite{gm} at total filling factor $\nu_{T} =1$. In the case of graphene bilayers excitonic condensation is expected to be richer due to additional Landau level degree of freedom.~\cite{excitoncondbg}  Exciton condensation, or equivalently spontaneous interlayer phase coherence, is likely to occur when any two Landau levels that are localized in different layers are degenerate.  This type of order is sometimes referred to as quantum Hall superfluidity and is a special case of quantum Hall ferromagnetism.  It should occur\cite{excitoncondbg} in bilayer graphene at $|\nu|=1,3$ when the electric field between layers is very small.

Few layer graphene systems stacked in different ways give rise to distinct electronic properties at relevant energy scales which introduces an entirely new class of 2DEG systems.
Recent theoretical work has indicated that except for very weak fields multilayer graphene~\cite{pacomultilayer,hongkimultilayer} systems can be classified as a new class of 2DEG systems from here on referred to as chiral 2DEGs (C2DEGs). C2DEG models provide an accurate description of the low-energy properties of few layer graphene systems with a variety of different stacking arrangements consistent with symmetries. Due to the unusual Landau quantization of the single particle Hamiltonian, IQHEs in C2DEGs are expected to be remarkably different from that of semiconducting 2DEGs. Two examples of this, which have already been discussed, are single layer and bilayer graphene which exhibit unusual IQHEs described by Dirac continuum models. At low-energies these systems represent the $J=1$ and $J=2$ instances of the C2DEGs~\cite{hongkimultilayer} model. Apart from these very recently several groups have reported experimental studies on ABA and ABC stacked trilayer graphene.~\cite{trilayerIQHE} ABC trilayer graphene systems are approximately described by a single $J=3$ C2DEG model.

\section{Low-energy theory of single- and multi-layer graphene}

Graphene is a two-dimensional material with carbon atoms arranged at the vertices of a
honeycomb lattice. A simple $\pi$-orbital tight-binding model for its
electronic structure\cite{wallace}
suggests correctly that graphene is a gapless semiconductor
with linear dispersion about two nodal
points~\cite{rmpgraphene,semenoff}
located at the inequivalent corners of a hexagonal  Brillouin Zone (BZ) ${\bf K}$ and ${\bf K'}$.
In an electrically neutral graphene sheet the nodal points
coincide with the Fermi level.  When carriers are present due to
doping or gating, the nodal points lie at the center of
nearly circular Fermi lines.

In the case of Bernal
stacked bilayer graphene, the low-energy effective theory also yields nodal
points at the BZ corners, but the dispersion is quadratic.~\cite{mccannfalko}
In this section we derive these low-energy theories for graphene and
bilayer graphene, emphasizing that they are special examples of
C2DEG models.~\cite{hongkimultilayer}

\subsection{Low-energy theory of graphene}
\begin{figure}[t]
\begin{center}
\includegraphics[width=3in,height=1.8in]{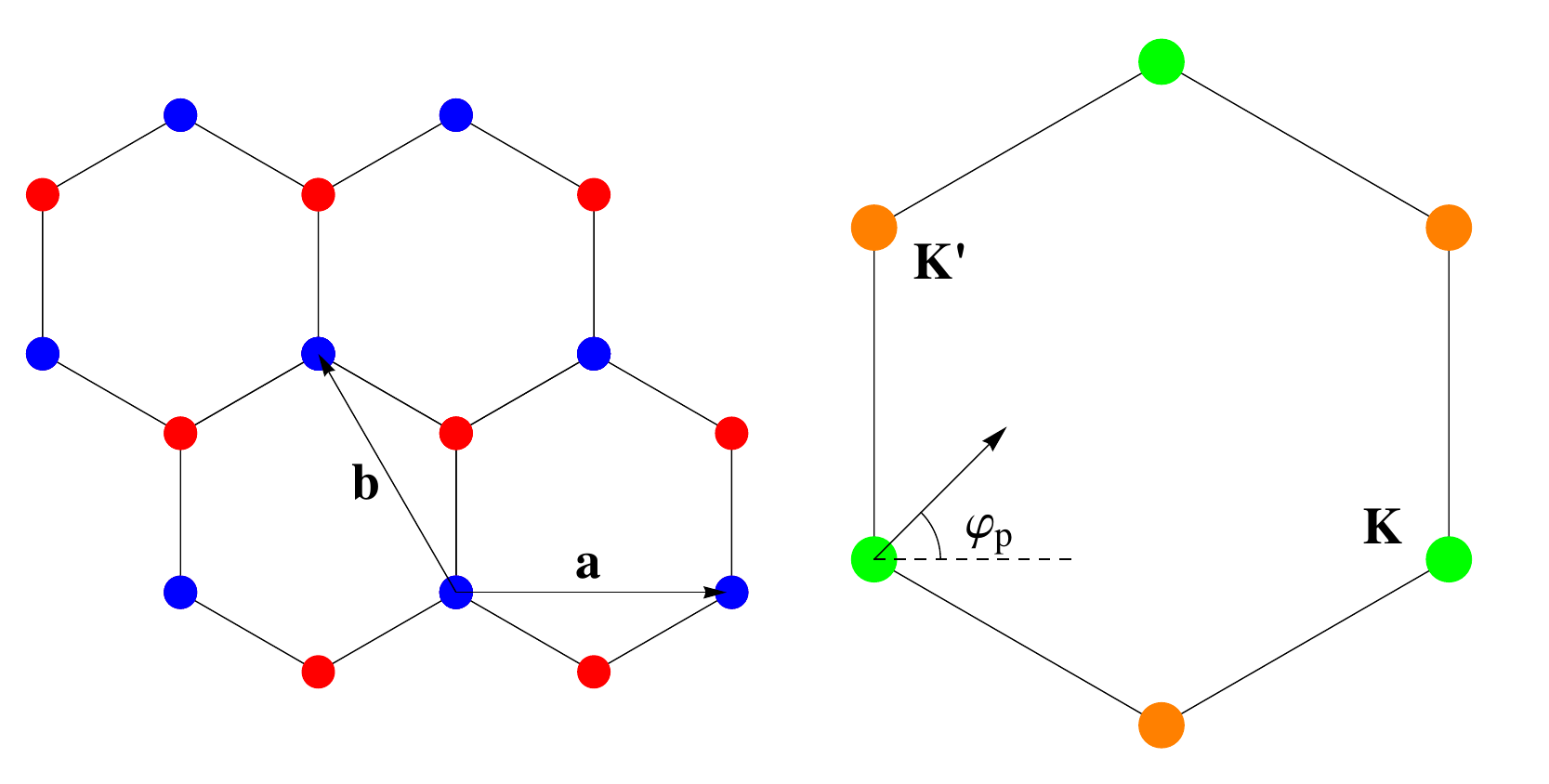}
\caption{Honeycomb lattice of a single layer graphite flake with one sublattice in red and
the other sublattice in blue. The lattice points form hexagonal plaquettes with the basis vector indicated by ${\bf a}$ and ${\bf b}$. In the continuum limit the sublattice degree of freedom may be regarded as a pseudospin. The reciprocal lattice Brillouin zone is also a hexagon with the Dirac points marked by ${\bf K}$ and ${\bf K'}$. When momentum ${\bf k}$ is measured from the Dirac point at the ${\bf K}$ or ${\bf K'}$ Brillouin zone corners, conduction and valence band eigenstates have their pseudospins oriented in the $\pm {\bf p}$ direction. This property is often referred to as chirality. The angle $\varphi_{\bf p}$ is the momentum-dependent phase difference between wavefunction amplitudes on the two sublattices. In bilayer graphene the pseudospin orientation angle is $2 \varphi_{\bf p}$.}
\label{figtwo}
\end{center}
\end{figure}
\indent
The honeycomb lattice is
a textbook example of a non-Bravias lattice, {\em i.e.} a lattice
with a basis of more than one atom per unit cell.~\cite{ashcroft}
It is illustrated in Fig.~\ref{figtwo} where A and B sublattice sites are
represented by red and blue circles.
In this section, starting from a $\pi$-band nearest neighbor tight-binding
description of electrons on a planar honeycomb lattice, we show that the low-energy states
of graphene are equivalent to those of massless Dirac fermions.
We choose the two basis vectors of a honeycomb lattice
to be ${\bf a} = a {\bf \hat{x}}$ and ${\bf b} = -a/2 {\bf \hat{x}} +\sqrt{3}/2 a {\bf \hat{y}}$,
where $a/\sqrt{3} = 0.142 nm$ is the carbon-carbon distance.

 Within the
nearest neighbor tight-binding approximation the Hamiltonian in the atomic site
basis can be written as
\begin{equation}
\mathcal{H}_{TB} = \left( \begin{array}{cc}
0 & -\gamma_0 f({\bf k})\\
-\gamma_0 f^{*}({\bf k}) & 0
\end{array} \right),
\end{equation}
where $\gamma_0 (\sim 2.8$eV) is the parameter
for nearest neighbor hopping between sublattices A and B and $f({\bf k})$ is defined as
\begin{equation}
f({\bf k}) = 1 + e^{i {\bf k} \cdot {\bf b}} + e^{i {\bf k} \cdot
({\bf b}+ {\bf a})}.
\end{equation}
The $\pi$ orbital site energies are identical by symmetry and chosen as the zero of energy.
The conduction and valence band energies  $E({\bf k}) = \pm \gamma_0 |f({\bf k})|$
both vanish at the two Brillouin-zone (BZ) corner points ${\bf K}$ and ${\bf K'}$.
These points are usually called Dirac points for reasons that will become clear towards the end of this section.
The Dirac point positions are given explicitly by ${\bf K} = -2\pi/(3a) {\bf \hat{x}} - 2\pi/(\sqrt{3}a) {\bf \hat{y}}$ and
${\bf K'}= -{\bf K}$. Expanding $f({\bf k})$ around the Dirac point ${\bf K}$ with
${\bf k} = {\bf K} + {\bf p}$
gives:
\begin{equation}
f({\bf p}) = {\bf p} \cdot \frac{\partial f}{\partial {\bf k}} |_{{\bf k}={\bf K}}=
\frac{\sqrt{3}}{2} \gamma_0 a (p_{x} - i p_{y}).
\end{equation}
The effective Hamiltonian for states that are close to the nodal  ${\bf K}$ point is therefore
\begin{equation}
\label{grapheneham}
\hat{\mathcal{H}} = v_{D}  {\bm \sigma} \cdot {\bf \hat{p}},
\end{equation}
where $\sigma_{i}$ are Pauli matrices that act on the sublattice
degree-of-freedom,  $\hat{p}_i = -i \hbar \partial_i $ is an
envelope function momentum operator.
Here $v_{D} = (\sqrt{3}/2) \gamma_0 a \approx 1
\times 10^{6} m/s $ is graphene's Fermi velocity which is approximately 1/300 times the speed of light.
The sublattice representation for valley ${\bf K}'$ can be chosen so that
the general Hamiltonian is $\mathcal{H}({\bf p}) = \xi v_{D}  {\bm \sigma} \cdot {\bf p}$ with $\xi = \pm 1$ for valleys ${\bf K}$ and ${\bf K'}$ respectively. To be consistent the pseudospinor for valley ${\bf K}$ is defined as $\psi^{\dagger}_{\bf K} = (\phi_{A,{\bf K}}^{\dagger}, \phi_{B,{\bf K}}^{\dagger})$ whereas the pseudospinor for valley ${\bf K'}$ is defined as $\psi^{\dagger}_{\bf K'} = (\phi_{B,{\bf K'}}^{\dagger}, \phi_{A,{\bf K'}}^{\dagger})$ and $\phi_{A}(\phi_{B})$ are envelop wavefunctions for the $A$ and $B$ sublattices measured from their respective valleys. So the sublattice labels should be reversed when interchanging the valleys. The band energy dispersion is linear near both Dirac points ${\bf K}$ and ${\bf K'}$: $\epsilon({\bf p}) = \pm v_{D}  |{\bf p}|$.

The low energy band model
for graphene, described above, is equivalent to
a model for two-dimensional massless Dirac fermions with a material specific speed
of light. When the Fermi energy lies at the Dirac points the system is neutral.
The Fermi energy lies above (below) the Dirac point for electron (hole)
doped graphene.
The eigenstates around valley ${\bf K}$ are:
\begin{equation}
\label{grapestates}
|\pm, {\bf p} \rangle_{\bf K} = \frac{1}{\sqrt{2}}
\left( \begin{array}{c}
 e^{ -i \varphi_{{\bf p}}/2}\\
\pm e^{ i \varphi_{{\bf p}}/2} \end{array} \right),
\end{equation}
where $\varphi_{\bf{p}} = \tan^{-1}(p_{y}/p_{x}) $ and $+$ labels the
conduction ($\bold{\pi}^{*}$)  band  case and $-$ the valence ($\bold{\pi}$)  band.
Note that the eigenstates  change sign if the phase $\varphi$ is rotated by $2\pi$.
This is the $\pi$ Berry phase mentioned earlier.

Chirality for pseudospin $1/2$ objects is defined as the projection of the pseudo(spin) operator in
the direction of momentum \cite{peskin}:
\begin{equation}
\hat{h} = \xi \frac{{\bm \sigma} \cdot {\bf \hat{p}}}{|{\bf p}| }.
\end{equation}
From this definition of the chirality operator it is easy to see that it commutes
with the Hamiltonian (\ref{grapheneham}), and
that $|\pm, {\bf p} \rangle_{\bf K}$ and $|\pm, {\bf p} \rangle_{\bf K'}$ are also eigenstates
of the chirality operator. Electrons (holes) have positive (negative) chirality for valley
${\bf K}$ ($\xi =1$).  Chirality is a good quantum number only as long as Hamiltonian (\ref{grapheneham}) is valid, and therefore holds only as an asymptotic property that is precise when close to the Dirac points. The eigenstates around ${\bf K}$ and ${\bf K'}$ are related by time reversal symmetry leading to opposite chirality in the two valleys. \\
\indent

\subsection{Low-energy theory of Bilayer graphene}
\begin{figure}[t]
\begin{center}
\includegraphics[width=3.0in,height=2.0in]{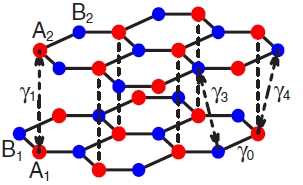}
\caption{Schematic illustration of the structure of bilayer graphene with two shifted honeycomb lattices. The direct hopping process between red sites in the bottom and top layer is represented by the hopping parameter $\gamma_{1}$. At the Dirac points this terms creates bonding and antibonding band eigenstates that are concentrated on the $A_{1}$ and $A_{2}$ orbitals. The low energy sites $B_{1}$ and $B_{2}$ are indicated in blue. The gapless conduction and valence bands have quadratic dispersions and are concentrated on these lattice sites. Direct hopping between the low-energy sites (the $\gamma_{3}$ process) leads to trigonal warping of the band dispersions.}
\label{figthree}
\end{center}
\end{figure}

\indent
The stacking arrangement of the two graphene layers in Bernal bilayers is illustrated in Fig~\ref{figthree}. Like single-layer graphene, Bernal stacked bilayer graphene has gaps that vanish at ${\bf K}$ and ${\bf K'}$ and a neutral system Fermi level that coincides with the nodal energies. In this section we start from the single-layer graphene low-energy Hamiltonian which has four component spinors corresponding to the four carbon sites per unit cell in this 2D crystal.
It is then possible to derive an effective two-component model which applies over a more
limited energy range. \\
\indent
The key to the bilayer graphene electronic structure is the spatial structure described in Fig.~\ref{figthree}. We label the sublattice sites as $A_{1}(B_{1})$ and $A_{2}(B_{2})$ in the bottom(top) layers respectively. Two of the four inequivalent atomic sites ($A_{1}$ and $A_{2}$) are stacked on top of each other, while the others ($B_{1}$ and $B_{2}$) do not have a near-neighbor in the other layer.  The $A_{1}-A_{2}$ interlayer hopping amplitude, $\gamma_{A_{1}-A_{2}} \equiv \gamma_{1} \approx 0.4eV$, leads to the formation of a dimer state of the $A_{1}-A_{2}$ orbitals. This process drives the energies of electrons on these sites away from the Fermi level, leaving one low-energy site per unit cell in each layer. An additional interlayer hopping process from $B_{1}$ to $B_{2}$, $\gamma_{B_{1}-B_{2}} \equiv \gamma_{3} << \gamma_{1}$ leads to a trigonal warping of the band structure~\cite{mccannfalko} at the lowest energies.  We neglect this and other remote weak interlayer hopping processes for the moment and will argue later that they can be neglected in the large magnetic field limit. \\
\indent
We follow McCann and Falko\cite{mccannfalko} by starting from a
continuum model bilayer band Hamiltonian which adds sublattice-dependent
interlayer hopping to decoupled 2D-Dirac bands:
\begin{equation}
\label{fourbandham}
\hat{\mathcal{H}}_{4 \times 4} = \left( \begin{array}{cccc}
0 & v_{D}  \hat{p}^{\dagger} & 0 & 0 \\
v_{D} \hat{p} & 0 & \gamma_{1} & 0\\
0 & \gamma_{1} & 0 & v_{D} \hat{p}^{\dagger}
\\ 0 & 0 & v_{D} \hat{p} & 0
\end{array} \right) ,
\end{equation}
for valley {\bf K} where $\hat{p} = \hat{p}_{x} + i \hat{p}_{y}$ and $v_{D}$ is graphene's Fermi velocity.
The above Hamiltonian operates on a four-component envelope function spinor wavefunction:
$\Phi^{\dagger} = (\psi^{\dagger}_{B_{1}, {\bf K}},\psi^{\dagger}_{A_{1}, {\bf K}}, \psi^{\dagger}_{A_{2}, {\bf K}},
\psi^{\dagger}_{B_{2}, {\bf K}}) $ where $\psi $ denotes the envelope function on the
corresponding atomic site. The energy spectrum $\epsilon^{\pm}_{\alpha}({\bf p}) $ takes the form
\begin{equation}
\label{relmassdisp}
\epsilon^{\pm}_{\alpha}({\bf p}) = \gamma_{1} \cos(\frac{\alpha \pi}{3}) \pm
\sqrt{v_{D}^{2}|{\bf p}|^{2} + \gamma_{1}^{2} \cos^{2}(\frac{\alpha \pi}{3})},
\end{equation}
with $\alpha = 1,2$.
As expected from our earlier discussion, two of the four bands have energies
$\epsilon^{+}_{1}({\bf p}) \approx \gamma_{1} $ and $\epsilon^{-}_{2}({\bf p}) \approx -\gamma_{1} $ and
are well separated from the Fermi level in systems with moderate carrier densities.
The other two bands are degenerate at the Fermi level and exhibit
quadratic dispersion for small ${\bf p}$. Expanding the low-energy bands $\epsilon^{-}_{1}({\bf p}) $ and $\epsilon^{+}_{2}({\bf p}) $ for $|\epsilon| \ll \gamma_1$ gives
\begin{equation}
\label{lowene}
\epsilon({\bf p}) = \pm \frac{p^{2}}{2 m},
\end{equation}
resembling the quadratic dispersion of a massive particle with an effective
mass $m = \gamma_{1}/2v_{D}^{2} \approx 0.054 m_{e}$, coincidentally a value that
is not very different from the effective mass in GaAs ($m^{*} = 0.064m_{e}$).  The range of validity of the two-band
bilayer graphene model can be estimated more precisely using Eq.(\ref{relmassdisp})
which implies the condition $|\epsilon| \ll \gamma_{1}/4$. We can therefore expect the two-band bilayer model, the $J=2$ C2DEG model as we explain below, to be accurate up to energy $\sim 100 {\rm meV}$.  The range of validity of the C2DEG model is therefore much more limited
in the bilayer than in the single-layer graphene case, for which it applies up to the eV energy scale.
The low-energy chiral behavior has already been observed in quantum Hall experiments on bilayer graphene.~\cite{bg_iqhe} We leave discussion of the Landau level spectrum to the next section. \\
\indent
The low-energy effective Hamiltonian for Bernal stacked bilayer graphene can be derived using
degenerate state perturbation theory.
The dimer states formed by vertical interlayer hopping between $A_{1}$
and $A_{2}$ form a high-energy space selected by the
projection operator $\hat{\mathcal{P}}_H = {\rm diag}(0,1,1,0)$. The low-energy subspace is spanned by sites $B_{1}$ and $B_{2}$ selected by the projection operator $\hat{\mathcal{P}}_L={\rm diag}(1,0,0,1)$.
We proceed by decomposing (\ref{fourbandham}) as:
\begin{equation}
\hat{\mathcal{H}} = \hat{\mathcal{H}}_0 + \hat{\mathcal{V}},
\end{equation}
where $\hat{\mathcal{H}}_0 = \hat{\mathcal{P}}_H \hat{\mathcal{H}} \hat{\mathcal{P}}_H $
and $\hat{\mathcal{V}} = \hat{\mathcal{H}} - \hat{\mathcal{H}}_0 $.
The first-order contribution in perturbation theory vanishes leaving us with a
second-order contribution to the low-energy effective Hamiltonian:
\begin{equation}
\hat{\mathcal{H}}^{(2)}_{eff} = \hat{\mathcal{P}}_L\hat{\mathcal{V}}\hat{\mathcal{P}}_H
\frac{1}{E-\hat{\mathcal{H}}_0} \hat{\mathcal{P}}_H \hat{\mathcal{V}}\hat{\mathcal{P}}_L,
\end{equation}
which for $E \ll \gamma_1$ gives
\begin{equation}
\label{bilayerham}
\hat{\mathcal{H}}^{(2)}_{eff} = - \frac{v_{D}^{2}}{\gamma_{1}} \left(
\begin{array}{cc} 0 & (\hat{p}^{\dagger})^{2}
\\ (\hat{p})^{2} & 0 \end{array}
\right).
\end{equation}
Here $\hat{\mathcal{H}}^{(2)}_{eff}$ acts on the spinor containing only the low-energy sites $\Psi^{\dagger} = (\psi^\dagger_{B_{1},{\bf K}},\psi^\dagger_{B_{2},{\bf K}})$.  Identifying $m = \gamma_{1}/2v_{D}^{2}$ gives the same quadratic dispersion
as before. The eigenstates for (\ref{bilayerham}) are obtained by
substituting $\varphi_{\bf{p}} \to 2 \varphi_{\bf{p}}$ in Eq. (\ref{grapestates}).

In Bernal stacked bilayer graphene the eigenstates exhibit a Berry's phase of $2 \pi$
while the conduction and valence bands exhibit gapless parabolic dispersion.
This 2DEG system is qualitatively distinct from both single-layer graphene and normal semiconducting 2DEGs.
In the next section we introduce effective Hamiltonians for C2DEGs,
which include the low-energy effective Hamiltonians of graphene and bilayer graphene as specific examples.
C2DEG models can be distinguished by a chirality index $J$.
Graphene and Bernal stacked bilayer graphene are members of this class with chirality indices
$J=1$ and $J=2$ respectively. In (\ref{bilayerham}) above the subscript $2$ can refer either
to the order of perturbation theory or to the chirality index $J=2$.\newline
\indent
Including the remote interlayer hopping $\gamma_{3}$ process and
allowing for the possibility of an onsite energy difference between layers $\Delta_{V}$
gives additional contributions to the effective Hamiltonian:~\cite{mccannfalko}
$\mathcal{H}^{(2)}_{eff} \to \mathcal{H}^{(2)}_{eff} + h_{w} + h_{\Delta}$ where:
\begin{eqnarray}
h_{w} + h_{\Delta} &=& v_{3} \;  \left( \begin{array}{cc}
0 & \hat{p} \\
\hat{p}^{\dagger} & 0 \end{array} \right) \\ \nonumber
&+&  \Delta_{V} \;
\left( \begin{array}{cc}
\frac{1}{2} - \frac{1}{2m \gamma_{1}} \hat{p}^{\dagger} \hat{p} & 0 \\
0 & -\frac{1}{2} + \frac{1}{2m \gamma_{1}} {\hat p} {\hat p}^{\dagger} \end{array}
\right),
\end{eqnarray}
where $v_3 = (\sqrt{3}/2) \gamma_3 a$.
$\gamma_{3}$ leads to a distortion of the band structure that is invariant under
$120^{\circ}$ rotations, trigonal warping, which does not play an essential role in the high magnetic field limit.
The external potential difference $\Delta_{V}$ between layers opens a gap in the spectrum and modifies the shape of the conduction and valence bands.
The modified dispersion has a Mexican hat shape~\cite{mexicanhat} in which the band extrema occur along circular lines.

The effect of interactions on undoped bilayer graphene's unusual band structure,
which we have neglected to this point,
is particularly interesting. Due to the absence of a Fermi surface, interaction effects
at the charge neutrality point can lead to non-Fermi liquid behavior with ill defined quasiparticle excitations
and a log squared behavior of the self-energy.~\cite{yafisandkun}
It has also been suggested that interactions can lead to the formation of gapped~\cite{Min,gapped2} or gapless~\cite{gapless} broken symmetry ground states. Recent experiments on suspended dual gated bilayer graphene samples do indicate anomalies near zero carrier density in bilayer graphene in the absence of a magnetic field\cite{Yacoby,Schonenberger,Novoselov,Lau} which we discuss further below.

\subsection{Two Dimensional Chiral Fermions}
\indent
Graphene and Bernal stacked bilayer graphene are unique and distinct two-dimensional materials
whose low-energy electronic properties are well described by chiral fermion models.\cite{hongkimultilayer}
The single particle chiral Hamiltonian with a chirality index $J$ is
\begin{equation}
\label{genchiralham}
\hat{\mathcal{H}}_{J} = \alpha_{J} \left(
\begin{array}{cc}
0 & \xi^J (\hat{p}^{\dagger})^{J}
\\ \xi^J (\hat{p})^{J} & 0 \end{array}
\right),
\end{equation}
where $\alpha_J$  has the dimensionality of
$ \epsilon/|{\bf p}|^J$, $\hat{p}= \hat{p}_x + i\hat{p}_y$ is momentum
measured from nodal points in the Brillouin-zone (BZ), and $\xi = \pm 1$ labels the two nodal points ${\bf K}$ and ${\bf K'}$ in the BZ. The pseudospin doublets in the respective valleys
${\bf K}(\xi =+1)$ and ${\bf K'}(\xi = -1)$ are defined as
$\Phi^{\dagger}_{\xi = +1} = (\phi^{\dagger}_{\uparrow},\phi^{\dagger}_{\downarrow})$ and
$\Phi^{\dagger}_{\xi = -1} = (\phi^{\dagger}_{\downarrow},\phi^{\dagger}_{\uparrow})$.
From the analyses of the earlier sections we can see that graphene and Bernal stacked bilayer graphene's effective Hamiltonians belong to this class with a chirality indices $J=1$ and $J=2$ respectively.
The higher $J$ models apply approximately to ABC stacked $J$-layer systems; although the
relative importance of corrections to the model increases with $J$ and the energy range over which it applies approximately tends to decrease with $J$. For general N-layer graphene systems, the Hamiltonian can be approximately decomposed at low energies into a set of decoupled C2DEG models with $\sum_i J_i = N$ and individual $J_i$ that depend on the
detailed stacking sequence.~\cite{hongkimultilayer} The energy dispersion of (\ref{genchiralham}) is $\epsilon_{J}({\bf p}) = \pm \alpha_J p^{J}$ with chiral band eigenstates:
\begin{equation}
\label{generalJestates}
|\pm, {\bf p} \rangle = \frac{1}{\sqrt{2}}
\left( \begin{array}{c}
 e^{i J \varphi_{\bf p}/2} \\
\pm e^{i J \varphi_{\bf p}/2}
\end{array} \right).
\end{equation}
Just as before the pseudospin spinors in the different valleys have opposite chirality. \\
\indent
These systems can be identified as different classes of zero-gap semiconductors. In each case the positive (negative) energies are identified with the conduction (valance) bands. The energy bands exhibit degenerate points ${\bf K}$ and ${\bf K'}$ in momentum space which are defined as
the points where the conduction and valence bands meet. In neutral structures
(i.e undoped with holes or electrons) the Fermi energy lies at the these degenerate points. \\
\indent
Chiral Hamiltonians satisfy time reversal and inversion symmetries.
The presence of two valleys is crucial for the time reversal invariance of chiral
Hamiltonians.  Time reversal symmetry for a chiral Hamiltonian refers to the
property that $(\Pi \otimes \sigma_{x})\mathcal{H}^{*}_{J}({\bf p}) (\Pi \otimes \sigma_{x}) =
\mathcal{H}_{J}(-{\bf p})$ where $\Pi$ swaps $\xi=+1$ and $\xi=-1 $ in valley
space. Chiral Hamiltonians also have the spatial inversion symmetry property
 $(\Pi \otimes \mathcal{I})\mathcal{H}_{J}({\bf p}) (\Pi \otimes \mathcal{I}) =
\mathcal{H}_{J}(-{\bf p})$. \\
\indent
The quasiparticles described by $\mathcal{H}_{J} $ acquire a Berry phase of $J \pi$
upon adiabatic propagation along a closed orbit. This has unusual consequences for the
single particle properties of chiral systems, the most notable of which is the
unusual Quantum Hall effects that have been measured in transport experiments in some of
these systems.\cite{graphene_qhe1,graphene_qhe2,bg_iqhe}
Graphene and Bernal stacked bilayer graphene  remain the primary focus of this review,
although some important properties have systematic $J$ dependences
which we shall mention next. \\

\section{The Integer Quantum Hall Effect of Chiral fermions}

In this section we evaluate the spectra of chiral fermion models
in the presence of a magnetic field, and discuss the
unusual IQHEs which follow from these spectra.
We also review transport experiments which verify the applicability of these
models to single layer and bilayer graphene.

\subsection{Chiral Fermions in a Magnetic Field}
\label{CHFmagfield}



%
In the presence of a uniform magnetic field applied in the direction
perpendicular to the plane of the C2DEG (${\bf B} = B \hat{z}$),
the momentum operator in (\ref{genchiralham}) is modified to
$\hat{\pi} = \hat{p} + (e/c) A$ where $A = A_x + iA_y $, and ${\bf A}$ is the vector
potential with ${\bf B} = {\bf \nabla} \times {\bf A}$.  We define the usual Landau level raising and lowering operators: $\hat{a}^{\dagger}$ and $\hat{a}$ with $\hat{a}^{\dagger} = (l_{B}/\sqrt{2} \hbar) \hat{\pi}$, where $l_{B} = (\hbar c/eB)^{1/2} = 25.6/(\sqrt{B[{\rm{Tesla}}]}) {\rm{nm}}$ is the magnetic length. Eq. (\ref{genchiralham}) for $\xi=+1$ then becomes :
\begin{equation}
\label{generalJmagfield}
\hat{\mathcal{H}}_{J} = \alpha_J \bigg( \frac{\sqrt{2} \hbar}{l_{B}} \bigg)^{J} \left(
\begin{array}{cc}
0 & \hat{a}^{J} \\
(\hat{a}^{\dagger})^{J} & 0
\end{array} \right).
\end{equation}\\
\indent
One immediate property of this Hamiltonian is the appearance of zero-energy eigenstates of the form $\Phi^{\dagger} = (0,\phi_{n})$ for
$n= 0,\cdots,J-1$.  Here the $\phi_{n}$ is the $n^{th}$ Landau level wavefunction for an ordinary 2DEG which is annihilated by $a^{k}$ for any $k > n$. Note that these zero-energy eigenstates are localized on the $\uparrow(\downarrow)$ pseudospin in the ${\bf K}({\bf K'})$ valley,
which for graphene corresponds to localization of these Landau levels on sublattice A (B) and for
bilayer graphene to localization on the top (bottom) layer. This property leads to a $4J$-fold degenerate (including valley and spin) zero-energy Landau level. This degeneracy is, of course, on top of the normal Landau level {\em one-state-per-flux-quantum} degeneracy.

The presence of wavefunctions with spatial structures
that appear at different energies in the ordinary non-relativistic 2DEG model in the same degenerate manifold
can create some terminological confusion.  We will refer to the wavefunctions $\phi_{n}$ as Landau level $n$ orbitals and
(as already anticipated) use upper case letter $N$ to distinguish levels with different Landau-quantized
band energies in the C2DEG model in a magnetic field. Thus the $N=0$ Landau level eigenstates of a C2DEG model includes states with orbital Landau level indices from $n = 0$ to $ n = J-1$.

The energies of higher $N$ Landau levels are given by:
\begin{equation}
\epsilon_{N,J} =  s \alpha_J \bigg( \frac{\sqrt{2} \hbar}{l_{B}} \bigg)^{J}
\sqrt{N(N-1)...(N-J+1)},
\end{equation}
where $s = \pm$ distinguishes positive energy conduction band ($s=+1$) and
negative energy valence band ($s=-1$) Landau levels.  The corresponding eigenstates are
\begin{equation}
\label{magfieldestate}
\Phi_{\xi=1} =\frac{1}{\sqrt{2}}
\left( \begin{array}{c}
s \phi_{n-J} \\
\phi_{n}
\end{array} \right),
\end{equation}
for valley ${\bf K}$. The eigenstates of these Landau levels are a linear superposition of 2DEG Landau level orbitals $n$ on one
sublattice and $n-J $ on the other sublattice.

The energy spectrum of a chiral system in a magnetic field is remarkably different from that of a normal 2DEG
in which the Landau levels are equally spaced with the gap between adjacent levels equal to
$\hbar \omega_{c} $ ($\omega_{c} = eB/mc$), and no zero energy eigenstates.
For a chirality $J$ system the level spacing
energy scale varies with field as $B^{J/2}$.  Spacings are larger near the Dirac point for $J=1$ and
smaller near the Dirac point for $J > 2$.  The unusual Landau level structure of single-layer and
bilayer graphene has been directly measured in scanning tunneling microscopy experiments on graphite
subjected to a perpendicular magnetic field.~\cite{andrei,stroscio}
The Landau level spectrum for chiral fermions leads to a family of anomalous IQHEs which have now
been verified experimentally in single layer, bilayer, and trilayer graphene. In next subsection we discuss
the unusual IQHE of chiral fermions explicitly.  \\

\subsection{Integer Quantum Hall Effect}
\begin{figure}[t]
\begin{center}
\includegraphics[width=3.2in,height=2.0in]{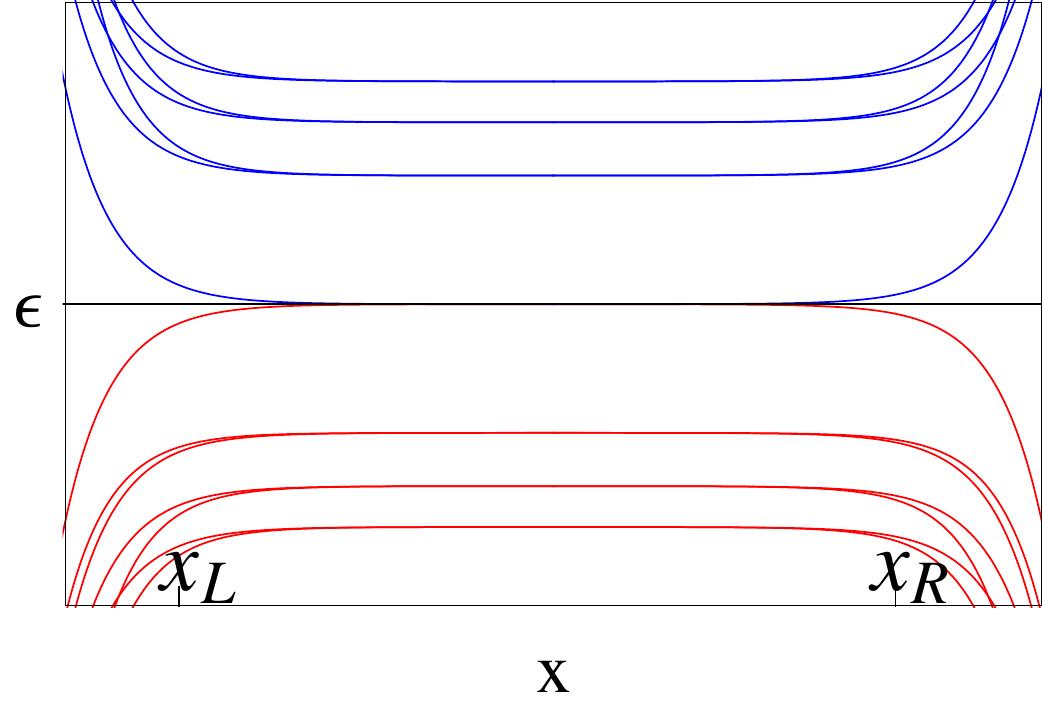}
\caption{Schematic representation of the energy dispersion of the edge states of a graphene strip in a magnetic field within the continuum model. The electron-like (hole-like) band dispersions are indicated in blue (red). The $|N|>0$ electron and hole-like bands are four-fold degenerate (spin and valley), whereas the $N=0$ bands are two-fold spin degenerate. $x_{L}$ and $x_{R}$ represent the respective left and right sample edges.}
\label{figfive}
\end{center}
\end{figure}

As explained in the introduction, the quantum Hall effect is a transport anomaly which
signals a jump in the chemical potential, a {\em charge gap}, at a magnetic field-dependent
density.  In translationally invariant continuum models the only length (and hence density)
scale is the magnetic length, which measures the area that encloses one quantum of magnetic flux. It follows that charge gaps in uniform 2DEG systems can occur only at fixed values of the
filling factor $\nu$, which is defined as the number of electrons per flux quantum.
(In graphene systems the electron density is measured relative to the density of a neutral sheet.)  A charge gap at filling factor $\nu$, when combined with incompressible state quasiparticle localization, leads to plateaus along which the longitudinal conductivity
$\sigma_{xx} \to 0 $ and the Hall conductivity $\sigma_{xy} \to \nu e^{2}/h $.  The use of the word {\em plateau} emphasizes that
because of charged excitation localization in the limit $T \to 0$ these values apply over a finite range of magnetic field, of carrier density, or of any other experimentally controllable parameter on which the 2DEG Hamiltonian depends. The relationship of the Hall conductivity to Chern numbers can be used to show that the quantum Hall effect can occur only at integer values of $\nu$ in the absence of electron-electron interactions.\\
\begin{figure}[t]
\begin{center}
\includegraphics[width=3.0in,height=2.4in]{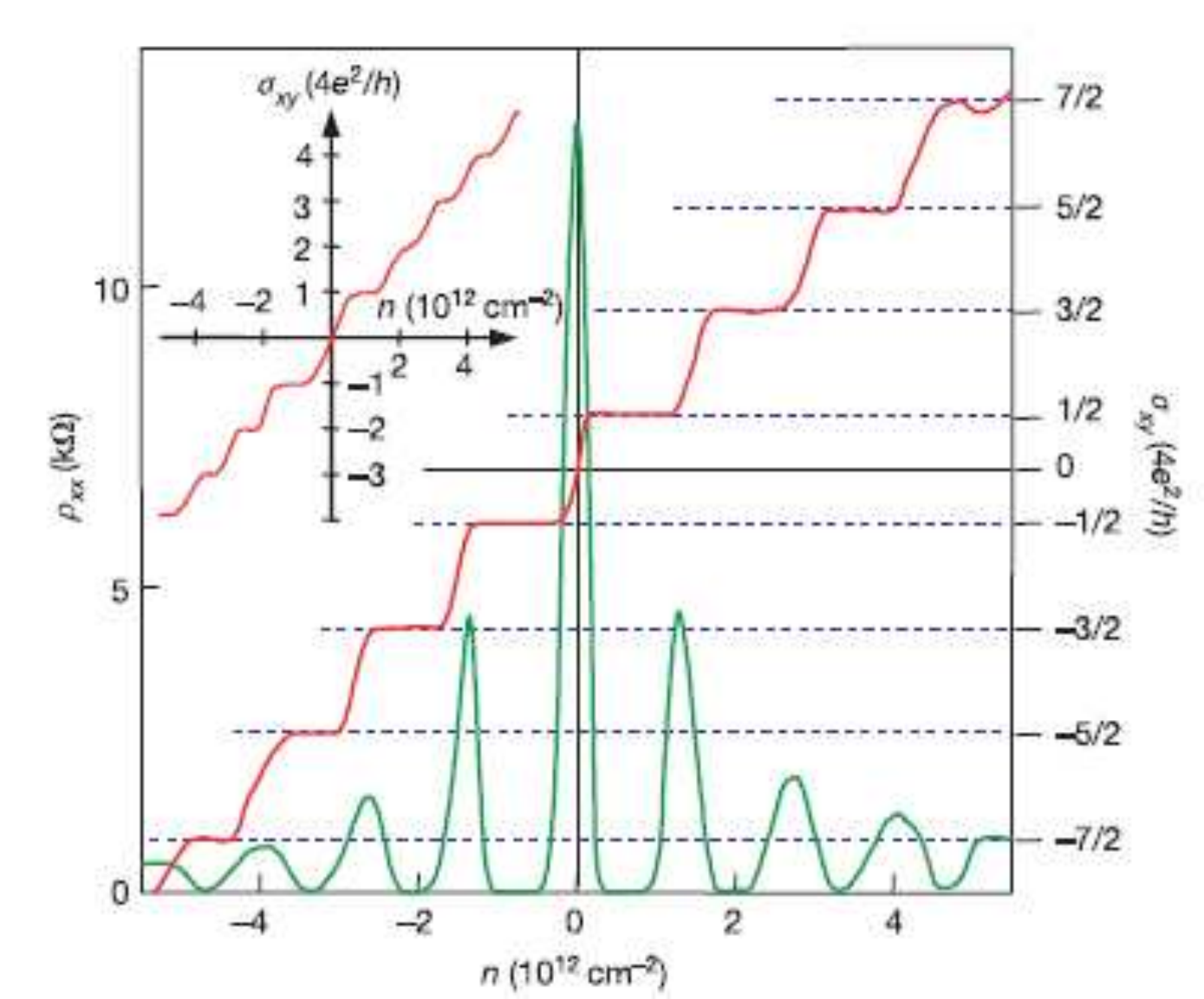}
\caption{(Adapted from Ref.\onlinecite{graphene_qhe1}(see also Ref.\onlinecite{bg_iqhe}) )  The Hall conductivity $\sigma_{xy}$ and the longitudinal resistivity $\rho_{xx}$ of monolayer graphene as a function of carrier density at $B = 14$ T and $T = 4$ K. This data
demonstrates the half-integer QHE with chiral index $J=1$ (graphene). The inset shows the Hall conductivity $\sigma_{xy}$ in bilayer graphene with a chirality index $J=2$.}
\label{figfour}
\end{center}
\end{figure}
Gaps at rational fractional values of $\nu$ (usually ones with odd denominators~\cite{sds}) have been studied for decades in ordinary 2DEGs and are now starting to be studied in graphene systems.  They can be produced only by
many-body electron-electron interaction effects. Incompressibility at a magnetic field dependent density always leads to gapless excitations localized at the edges of the system, the smoking gun signature of the QHE.~\cite{allanqhreview}\\
\indent
In C2DEGs, as in any other system, incompressibility
at a magnetic field dependent density implies Hall quantization.
The filling factors at which gaps are produced by Landau-level formation
alone differ however.  When the Zeeman energy and $\Delta_{V}$ are set to $0$,
gaps appear at the following set of integer filling factors (as indicated in Fig. \ref{figfour}):
\begin{equation}
\label{quantcond}
\nu = \pm \; 4 \;  (N + \frac{J}{2}), \qquad N = 0,1,2,\cdots.
\end{equation}
Note in particular that there is no gap at zero carrier density $\nu=0$.
The above quantization condition follows directly from the  energy levels of chiral
fermions in the presence of a perpendicular magnetic field discussed above.\\
\indent
To understand the edge state implications of this
unusual quantization condition let us first review the IQHE in a normal
2DEG in which Landau level quantization leads to the
energies $\epsilon_{n} = (n+1/2) \hbar \omega_{c}$.
Landau levels are equally spaced with spacing equal to the cyclotron energy $\hbar \omega_{c}$.
In the presence of disorder Landau levels get broadened. When the Fermi energy lies in a gap between
Landau levels, the only states at the Fermi energy when disorder is weak are chiral edge states which can carry current without dissipation, causing the longitudinal resistance to fall to zero.\cite{halperinedge} When $n$ Landau levels are filled, each Landau level gives rise to a gapless edge state branch. Because these $n$ channels carry current in parallel, $\sigma_{xy} = n e^{2}/h $.  Each time the chemical potential crosses a Landau level there is an extra contribution from a new gapless edge state channel and the Hall conductivity jumps by $e^2/h$.\\
\indent
The unusual edge state structure and edge transport unique to chiral fermions can be understood from the
Landau level spectrum and, in particular, from the presence of zero-energy eigenstates.~\cite{breyfertigedge}
We can identify the Landau levels of a chirality $J$ C2DEG as electron-like for $N > 0$ and hole-like for  $N < 0$ as indicated in Fig. \ref{figfive} for graphene ($J=1$). The $4J$-fold degenerate $N=0$ zero-energy state is neither electron-like nor hole-like. Given the spectrum of Landau levels for chiral fermions, the Hall conductance $\sigma_{xy}$ must jump by $4 J e^{2}/h$ when the chemical potential crosses the $N = 0$ Landau level.
The $4 \times J/2$ term in Eq. (\ref{quantcond}) accounts for the presence of $4J$ zero-energy Landau levels which give rise to $4J$ edge state branches.
The unusual property of the $N=0$ Landau level is that, as what one might guess on the basis of particle-hole symmetry,
half of its edge branches appear at energies above the bulk Landau level and half at energies below.
 The Hall conductivity on each plateau is therefore equal to the number of edge state branches in the corresponding gap.\\
\indent
Soon after the isolation and identification of graphene this unusual IQHE was measured
in Hall conductivity experiments in both single layer~\cite{graphene_qhe1,graphene_qhe2}
and bilayer graphene~\cite{bg_iqhe}. Fig.~\ref{figfour} shows the experimentally measured Hall
conductivity for monolayer and Bernal bilayer graphene.~\cite{bg_iqhe} For single-layer graphene
the gap between $N=0$ and $|N|=1$ Landau
levels is around 1300 K at 10 T compared to around 100 K in an ordinary 2DEG.
This is the reason the quantum Hall effect can be observed at room temperature,~\cite{roomtemphall}
a great surprise given historical norms.\\
\indent
Experiments now clearly show that at higher fields and in higher quality samples additional integer 
QH plateaus appear for both single layer~\cite{graphene_qhfexp,graphene_qhf2} and bilayer graphene~\cite{bg_qhf1,bg_qhf2}
that are not predicted in the Integer Quantum Hall sequence (see Eq.~\ref{quantcond}).
These new plateaus cannot be explained by Landau quantization alone. They must originate from the effect of Coulomb interactions or from explicit symmetry breaking one-body terms in the Hamiltonian - for example in the case of bilayer graphene from a large electric potential difference between the two layers.  Zeeman splitting is not on its own
strong enough to produce additional quantum Hall effects in samples with typical disorder strengths,
and must be enhanced by interactions. These additional gaps at integer filling factors are normally due to interaction-driven broken symmetries and have been anticipated based on both Quantum Hall Ferromagnetism~\cite{nomura,gmd,aliceafisher,ydm,abanin} and Magnetic Catalysis~\cite{gmss1,herbut,gmss2} scenarios, as discussed earlier. In the next section we discuss these additional integer plateaus at greater length.

\section{Interaction induced Integer Quantum Hall States}

\indent
In semiconductor 2DEG systems a wide variety of quantum Hall states
have been discovered which are driven by interactions,
{\em i.e.} incompressibilities develop
at filling factors $\nu$ at which the band Hamiltonian does not have a gap.
From a fundamental physics point of view the most important of these are
the ones that appear when a Landau level is fractionally filled, since they
lead to fractionally
charged excitations with fractional (and in some cases non-Abelian) statistics.
Another important class of examples in semiconductor 2DEGs
are the gaps which appear at odd integer filling factors.
These are often found experimentally to be much larger than the
Zeeman energy gap of the single-particle Hamiltonian.  In fact
gaps do appear at odd integer values of $\nu$ even when the Zeeman
energy vanishes.\cite{potemski}   The experimentally measured gaps are
usually essentially pure exchange gaps due to electron-electron interactions and
spontaneous spin polarization, {\em i.e.} due to what is often referred to as
quantum Hall ferromagnetism.  The notion of quantum Hall ferromagnetism has been
generalized to other broken symmetry induced gaps in the quantum Hall regime.\cite{jungwirthqhf} A particularly interesting aspect of quantum Hall ferromagnetism in semiconductors is the presence of {\em skyrmion} topological spin textures that carry electrical charge.\cite{sondhi,gm} \\
\indent
Band energy quantization for C2DEGs in the presence of  magnetic field
leads to a quantum Hall sequence with $\nu = \pm 4(N+J/2) $ with $\nu = \pm 2, \pm 6, \pm 10,\cdots$
for single layer and  $\nu = \pm 4, \pm 8, \pm 12,\cdots$ for bilayer graphene.
The four-fold degeneracy evident in the measured Hall conductivity of graphene is due to spin and valley degeneracy.~\cite{gusynin,graphene_qhe1,graphene_qhe2} Bilayers exhibit an eight-fold degeneracy for $N=0$ {\em only}
which is due to the additional orbital degeneracy between orbitals with $n=0$ and $n=1$ character.~\cite{mccannfalko,bg_iqhe}
When the spin Zeeman effect is included gaps appear at all even integer filling factors for single and bilayer
graphene, except in the bilayer case for $\nu = \pm 2 $. It was predicted theoretically that gaps would appear at all
integer filling factors in samples with disorder energy scales that are weaker than electron-electron interaction
scales in both single layer~\cite{nomura} and bilayer~\cite{ourbilayerqhf} graphene. This prediction has now been
verified for single layer graphene~\cite{graphene_qhfexp,graphene_qhf2,BNsubstrate} and also in bilayer
graphene~\cite{bg_qhf1,bg_qhf2}. In this section we discuss the physics of
interaction driven integer Quantum Hall states in graphene and in bilayer graphene.\\
\indent
Single layer graphene gaps for  $\nu \ne \pm 4(N+1/2)$ can be due either to explicit symmetry breaking
by one-body terms not included in our earlier analysis (that are usually enhanced
by interaction dressing), or to purely spontaneous symmetry breaking
by interactions.  Interaction driven additional integer quantum Hall states
have been anticipated by {\em Quantum Hall Ferromagnetism}~\cite{nomura,gmd,aliceafisher,ydm,abanin} (QHF)
theories which focus on interactions between states in the same Landau level, and by
{\em Magnetic Catalysis}~\cite{gmss1,herbut,gmss2} (MC) theories
which focus on the $\nu=0$ case  and assume that Landau level mixing
plays an essential role. These theories can be further distinguished by their experimental predictions.
While QHF theories predict that charge gaps and hence quantum Hall effects should be present at all
integer filling factors, MC theories predict that extra charge gaps should
appear at $\nu=0$. The recent experimental observation of QH plateaus at all integer filling factors~\cite{BNsubstrate}
for graphene on h-BN seems to support QHF theories. \\
\indent
Bilayer graphene also exhibits spontaneously broken symmetry states due to the presence of electron-electron interactions. In bilayer graphene QHF theories also predict ordered states which exhibits incompressibilities at all integers plateaus. The $N \neq 0$ broken symmetry states are expected to be characterized by spontaneous
spin and/or valley polarization. Because the $N=0$ Landau level includes an extra degree of freedom associated with the $n=0$ and $n=1$ orbital degree of freedom~\cite{mccannfalko} the physics associated with its polarization, which is expected
at odd total filling factors, is more interesting.  These orbitally broken symmetry states have
unusual Goldstone collective modes~\cite{ourbilayerqhf} associated with the dipoles induced by
inter-orbital coherence.  Apart from this the parabolic dispersion of bilayer graphene already makes it susceptible to interaction-induced symmetry breaking even at zero breaking magnetic field.~\cite{Min,gapped2,gapless} Experiments on doubly gated suspended bilayer graphene samples indicate that incompressibilities exist for vanishing magnetic fields,~\cite{Yacoby} although the nature of the associated broken symmetry has not been unequivocally\cite{Novoselov,Lau} established as of this writing.

\subsection{Interaction induced integer QH states in single layer graphene}

\subsubsection{Quantum Hall Ferromagnetism in Graphene}

Soon after the initial verification of the unusual IQHE of graphene,
experiments~\cite{graphene_qhfexp} at higher fields ($B > 20 T$) revealed
further plateaus $\nu = 0,\pm 1 $ and $\nu = \pm 4 $ that do not belong to
the IQH sequence of graphene. These additional states suggest
that the four-fold spin and valley degeneracy is lifted in a variety of
filling factor-dependent broken symmetry states.
Early tilted field experiments appeared to suggest that the states at $\nu = 0, \pm 4 $
are spin polarized, implying that spin degeneracy is broken at these filling factors,
whereas the $\nu = \pm 1$ states depend only on the perpendicular component
of the magnetic field, indicating that their gap is not spin related and corresponds
instead to broken valley degeneracy.~\cite{graphene_qhf2} However, more recent experiments indicate that the $\nu=0$ gap is not spin-polarized and that the $\nu = \pm 1 $ gap increases with in plane field.~\cite{andreadiscuss} Temperature dependent activated measurements indicate that the gaps at $\nu = 0, \pm 4$  are larger than those expected from Zeeman splitting alone. If these gaps are spin-splittings, they are apparently strongly enhanced by many-body effects. Recent studies of graphene on h-BN substrates, which are much more ordered than SiO$_2$ substrates, appear to indicate that a number of the gaps in higher Landau levels involve both spin and valley degrees of freedom.~\cite{andreadiscuss}\\
\indent
Theories of quantum Hall ferromagnetism focus on interactions within a Landau level and
the broken symmetry states they can induce. The discussion in the following
paragraphs follows this line of thought and neglects Landau-level mixing effects.
In these theories it is therefore convenient to introduce a reduced filling factor for each Landau level $N$:
\begin{equation}
\nu_{N} = \frac{N_{e}}{N_{\phi}} \leq 4,
\end{equation}
where $N_{e}$ is the number of electrons in the $N^{th}$ Landau level and $N_{\phi}$ is
the number of flux quanta enclosed in the system.
In the continuum theory of graphene the spin and valley degrees of freedom are equivalent flavor labels. When Zeeman coupling and other weak perturbations are neglected, the Hamiltonian is SU(4) invariant. At integer filling factors Quantum Hall ferromagnet ground states spontaneously break the SU(4) symmetry, ensuring good exchange energy for flavor indices within a Landau level and inducing a charge gap for excitations.\\
\indent
The integer filling factor ground state of the SU(4) invariant Hamiltonian is known
exactly.  For filling factor $\nu_{N}$
\begin{equation}
\label{wavefunction}
|\Psi_{0} \rangle = \prod_{1 \leq \sigma \leq \nu_{N}} \prod_{k} c^{\dagger}_{k,\sigma}
|0 \rangle,
\end{equation}
is an exact eigenstate of any SU(4) invariant electron-electron interaction Hamiltonian.
In (\ref{wavefunction}) $c^{\dagger}_{k,\sigma}$ is the electron creation operator
for intra-Landau level orbital index $k$ and flavor (valley/spin) label $\sigma $.
$|0 \rangle$ corresponds to the vacuum state in which Landau level $N$ is completely
empty and all lower Landau levels are completely filled. The wavefunction $|\Psi_{0} \rangle$ has
lower energy because it puts many electrons into the same
pseudospin state, minimizing their exchange energy to the maximum degree
allowed by the Pauli exclusion principle at filling factor $\nu_{N}$.

These states are exact eigenstates of
SU(4) invariant Hamiltonians at the integer partial filling factors $\nu_{N} = 0,1,2,3$,
but are they the ground states?
Quantum Hall ferromagnetism theories are based on the notion that they normally are.
Whether or not this idea is correct likely depends to some degree on the character of the
Hamiltonian, with Hamiltonians that are more strongly repulsive at short distances most
likely to support these ground states.  In the case of SU(2) (spin flavor only), it
seems clear from finite-size exact diagonalization calculations\cite{SU2ed} that they
are ground states for small $n$ Landau levels.  Exact diagonalization calculations for the
graphene SU(4) case also support the quantum Hall ferromagnetism
{\em ansatz} for small $N$ in graphene. Although it is not possible to prove rigorously
that the integer $\nu_{N}$ ground states have this form, even
when Landau level mixing is neglected, the case for quantum Hall ferromagnetism
at small $|N|$ appears to be quite strong.  Its neglect of Landau-level mixing effects
also appears to be soundly based although this approximation does not become
better, as it does in ordinary 2DEGs, as the field strength is increased.

When the quantum Hall ferromagnetism {\em ansatz} is correct, an
electron that is added to an integer partial filling factor state
will have to occupy a flavor Landau level that is empty, losing exchange energy in the process.
The charge gap is due to this loss of exchange energy. This incompressibility at a fixed
filling factor
leads to an integer quantized Hall conductivity and to vanishing longitudinal conductivity,
in accord with well-established quantum Hall phenomenology.\\
\indent
Since SU(4) is a continuous symmetry, when it is broken spontaneously the
systems must support gapless collective modes. These neutral,
spin-wave like modes have been studied,\cite{aliceafisher,ydm}
and their full spectra have been determined exactly.\cite{ydm}
Because the symmetry group is larger
than in the SU(2) case,\cite{gm} the number of spin-wave-like modes is larger than one.
The number of modes is given by $\nu_N(4-\nu_N)$ --- four modes
for $\nu_{N}=2$ and three modes $\nu_{N}=1,3$.\cite{ydm}
\indent
Perhaps more interesting is the appearance of topological solitons
in the order parameter called Skyrmions as low-energy charged
excitations of the systems.\cite{sondhi,gm} Again due to the larger
symmetry, Skyrmions come in more species in the SU(4) quantum Hall ferromagnets\cite{ydm}
case compared to the SU(2) quantum Hall ferromagnets studied earlier.\cite{sondhi,gm}

There is an important quantitative difference compared to the ordinary
2DEG case in the spatial structure of
electron-electron interactions that is due to the Dirac nature of
graphene's charge carriers.  In a magnetic field, the
two sublattice components of graphene's  Dirac-model
eigenspinors have {\em different} orbital Landau level wave functions
\cite{nomura} (see also Section \ref{CHFmagfield}).
As a result, the form
factor which determines the effective electron-electron interaction in a given
Landau level is quite different for Dirac fermions.\cite{nomura,gmd} One of
the consequences of this is that Skyrmions are the lowest energy charged
excitations not only in the lowest Landau level (which is the only stable Skyrmion case
for non-relativistic electrons\cite{wu}), but also in
some of the higher Landau levels\cite{aliceafisher,ydm}; more specifically they are
lowest energy charge excitations for $|N |\le 3$.\cite{ydm} These
high Landau level skyrmions have been observed in recent numerical
work.\cite{toke} \\
\indent
Because states in different valleys are certain to be on different sublattices in the $N=0$ case,
valley and sublattice (layer in the bilayer case) labels are equivalent when Landau level mixing is neglected.
This situation changes when mixing becomes important.  The role of Landau level mixing
can be increased in the single-layer case by increasing the strength of carrier-carrier interactions,
{\em i.e} by decreasing dielectric screening.  (This contrasts with the bilayer case in which
Landau level mixing can be enhanced simply by decreasing the
magnetic field strength.)  When mixing is important there is a qualitative difference between
sublattice and valley labels because sublattice is not a good quantum number.
When the valence Landau level breaks symmetry by establishing a condensate
which favors sublattice polarization, the wavefunctions of Landau levels far from the
Fermi surface are altered in a way which reinforces the condensate.
In the language of the 2D Dirac equation, a sublattice
polarization term is a mass term, as discussed further below.   Depending on the relative signs of the masses
for different spins and valleys, sublattice symmetry breaking leads to various kinds\cite{fanzhangpaper} of
density wave states and to gaps at a variety of integer filling factors that persist to
weak fields.  Quantum Hall ferromagnet states in which the
broken symmetry condensate produces a mean field term which favors occupation
of one sublattice over another and plays the role of this
mass in the 2D Dirac equation.  For total Landau level filling factor $\nu=0$
this subset of quantum Hall ferromagnet broken symmetry
states is related to the idea of magnetic catalysis.
Indeed there is a subset of quantum Hall ferromagnet states which, when
stable, tend to remain robust at weaker magnetic fields.  There is some evidence that the
$\nu=0$ state in both single-layer and bilayer graphene may have the type of order
anticipated by the magnetic catalysis scenario.

\indent
\subsubsection{Explicit SU(4) Breaking Interactions}
A number of small perturbations that explicitly break SU(4) symmetry are
present in graphene.  They can fix the orientation of the SU(4) order
parameter, {\em i.e.} select specific preferred flavor orbitals to occupy at
a particular partial filling factor,  and will also open small gap(s) in
some of the the collective mode dispersions.
In the following we discuss some of these perturbations. \\
\indent
{\em Zeeman Splitting:}--- This is probably the most obvious explicit
symmetry breaking interaction.  It lifts the spin degeneracy without affecting the valley degeneracy;
formally it reduces the SU(4) symmetry to a (valley)  SU(2) $\otimes$ U(1) symmetry.
Its influence is particularly important when $\nu_N=2$, in which case
it uniquely selects a single ground state which is a valley
singlet but has spins in the valence Landau level fully polarized
along the field direction.  In the absence of interactions the charge gap
associated with this state is precisely the Zeeman splitting, and in some
theories (to be discussed later) this is thought to be the sole
source of gap at $\nu=\pm 4$, when the $ N=1$ Landau level is half-filled. Quantitatively, the Zeeman splitting
is of order 10 K at 10 Tesla, compared to the electron-electron interaction scale
$e^2/\epsilon\ell \sim 500 K$ ($\ell=\sqrt{\hbar c/eB}$ is the
magnetic length); it thus qualifies as a small symmetry breaking
perturbation. \\
\indent
{\em Lattice Effects:}--- SU(4) symmetry is exact in the continuum
limit $a \ll \ell$, where $a$ is the lattice spacing. Lattice effects
break SU(2) symmetry
within the valley subspace.\cite{gmd,aliceafisher} The strength of
this symmetry breaking is proportional to $a/\ell$. For the lowest Landau level
($N=0$), lattice effects introduce an Ising anisotropy in the valley SU(2)
space,\cite{aliceafisher} while for $|N|>0$ an easy-plane (or XY)
anisotropy is introduced.\cite{gmd,aliceafisher}  Lattice Hartree-Fock
calculations\cite{jeiljungpapers} point to one important role of the lattice.
In a continuum model electrons do not change their
valley index when they scatter off another electron.  As a consequence
exchange interactions occur only within valleys.  Although the
neglected inter-layer exchange interactions are weak, they do
favor sublattice polarization states in which the sense of layer
polarization is the same for both valleys.  This inter-valley exchange
effect can be important in determining the character of $\nu=0$
quantum Hall states as we discuss further below.\\
\indent
{\em Disorder:}--- Although disorder potentials that are smooth on
atomic length scales do not break any
symmetry, they do lift Landau level degeneracy.  Disorder favors SU(4) singlet
ground states because more electrons can be placed where the
disorder potential is lowest when the four valley
degrees of freedom are exploited symmetrically.  In Ref. \onlinecite{nomura},
in which the notion of QHF was first introduced in the graphene context, a Stoner-like
criterion was derived for the competition between interactions and
disorder in quantum Hal ferromagnetism.
The situation was also analyzed in Ref. \onlinecite{aliceafisher},
in which the authors came to the conclusion that the samples
available at that time might not be clean enough for QHF, and the systems under study
at that time, and perhaps now, might actually be SU(4) paramagnets.  They then argued
that the combination of Zeeman and
symmetry breaking interactions can give rise to new QH states in
the paramagnetic regime.\\
\indent
On the other hand Abanin {\em et al.}\cite{abanin} pointed out an
interesting symmetry breaking effect of the disorder potential, which
may {\em favor} ordering: a random potential produced by strain in the
graphene lattice locally breaks the valley degeneracy in a random
manner; it thus acts like a random field along the $\hat{z}$ direction
in the valley subspace. This could force the valley SU(2) order parameter
into the XY plane, and allows for algebraic long-range order at {\em
finite} temperature and a Kosterlitz-Thouless transition.\\
\indent

\subsubsection{Magnetic Catalysis}

One set of theories~\cite{gmss1,gmss2,ezawa,herbut} related to interaction induced gaps in graphene's quantum Hall regime
focuses on interaction induced mass terms in the two-dimensional
Dirac equation that describes graphene electrons.  Interaction-induced masses
in the absence of a magnetic field are partly analogous to interaction-induced masses in elementary
particle physics.  {\em Magnetic catalysis} refers to the reduction in the interaction strength necessary
for spontaneous mass that occurs when a magnetic field is present.\\
\indent
Since a Dirac equation mass corresponds to a $\sigma_{z}$ term in the sublattice-pseudospin
mean-field Hamiltonian, it is always generated in states that have
$A$-$B$ sublattice polarization within each valley.
The broken symmetry states anticipated in magnetic
catalysis theories are ones in which the same sense of sublattice polarization occurs
in each valley.  From a physical point of view these are therefore density-wave states in which the honeycomb
lattice symmetry is reduced. Spontaneous mass generation in monolayer
graphene was in fact discussed as a theoretical possibility even before the experimental realization of the material.\cite{khvesh} The
excitonic gap imagined in this early work was attributed to an instability towards electron-hole
pairing in the presence of a constant magnetic field in $(2+1)$D-QED.
Magnetic catalysis theories are distinguished phenomenologically from QHF theories by their prediction that
spontaneously generated gaps should appear in the $N=0$ Landau level {\em only}.
Induced mass type order opens gaps between the conduction and valence bands, instead of
opening them within all Landau levels. When Zeeman splitting is included, these theories predict plateaus in
the Hall conductivity $\sigma_{xy}$ at $\nu=0,\pm 1$, and $\nu = \pm 2k$ for
$k=1,2,3,\cdots $. In Ref. \onlinecite{herbut}, it was argued that the
ordering that generates the mass could be either a charge density wave (CDW) or
a spin density wave (SDW), depending on the relative strength of
on-site and nearest neighbor electron-electron repulsion.  SDW states
have the opposite sense of sublattice polarization for opposite spins.\\
\indent
Earlier experimental results\cite{graphene_qhfexp} appeared to be consistent with the magnetic catalysis scenario since
the first odd integer QH states to be observed appeared at  $\nu=\pm 1$ in the $N=0$ Landau level.
(Assuming that spins are maximally polarized because of the Zeeman
field, gaps at $\nu=\pm 1$ require that valley or sublattice symmetry be broken,
consistent with the magnetic catalysis scenario.) However the recent observation of integer QH states at
$\nu = 3$ in suspended graphene samples and at all integer filling factors
for graphene on h-BN substrate seem to indicate the QHF scenario better describes the
physics which takes place in graphene sheets. In fact magnetic catalysis should be viewed
as a special case of quantum Hall ferromagnetism in which the broken symmetry state is a
charge-density-wave or spin-density-wave state. \\
\indent
CDW and SDW states are likely strong-interaction-limit broken symmetry states for
honeycomb lattice models in the absence of a magnetic field. If these broken
symmetries were incipient at $B=0$, a weak magnetic field would in all likelihood push
them over the edge --- hence magnetic catalysis.  If on the other hand $B=0$ graphene
systems are not close to instabilities, there is no reason to expect that
CDW and SDW instabilities would dominate in the presence of a magnetic field.
The experimental observation of QHE's at nearly all integer filling factors suggests that
it is Landau-level formation and interactions within groups of degenerate Landau levels,
the quantum Hall ferromagnetism scenario, that plays the essential role in
establishing these interaction-induced gaps.  If so, this piece of graphene
phenomenology suggests that the $B=0$ graphene system is not especially close to
an interaction-induced instability. However in bilayer graphene it appears that the story is different.
Some quantum Hall effects appear to persist to very weak magnetic fields\cite{Yacoby,Schonenberger,Novoselov,Lau}
and sometimes to $B=0$, with interactions playing an essential role. It appears likely that neutral bilayer
graphene band structure is unstable at $B=0$ against {\em weak} interactions, as predicted
theoretically\cite{Min,gapped2,gapless} (see section V. B).

\subsubsection{$\nu =0$ Quantum Hall State Transport Properties}
\indent
The $\nu =0 $ QH state in graphene has garnered particular interest since it does not have an
analogue in semiconducting 2DEGs with parabolic dispersion. The new physics is due to the relativistic Dirac dispersion for quasiparticles in graphene
which lead to conduction and valence bands (see Section III A) that are not separated by a gap.
Initial experimental studies of this state indicated that the transport phenomenology on the $\nu =0$ QH state is different from that of the other integers. This unusual behavior is evident both in $\rho_{xx}$ and
$\rho_{xy}$, the diagonal and off-diagonal resistivity respectively. The experimentally measured  Hall resistivity does not display either a clear quantized plateau or a zero resistance state;\cite{abaningeim} however a step-like feature was evident in the off-diagonal conductivity at $\sigma_{xy} =0 $. The longitudinal resistivity did not exhibit a deep minimum normally characteristic of the conventional QHE. Later experiments on the $\nu =0$ state reported a steeply divergent Dirac point resistance at low temperatures.~\cite{graphene_ong} The authors argued that this insulating behavior followed Kosterlitz-Thouless (KT) type scaling with respect to the magnetic field and, was interpreted as a quantum-KT phase transition by assuming a valley split XY state. This field induced insulating behavior has also been observed in suspended graphene samples in a two terminal setup.~\cite{graphene_fqhe1,graphene_fqhe2} \\
\begin{figure}[t]
\begin{center}
\includegraphics[width=3.2in,height=2.0in]{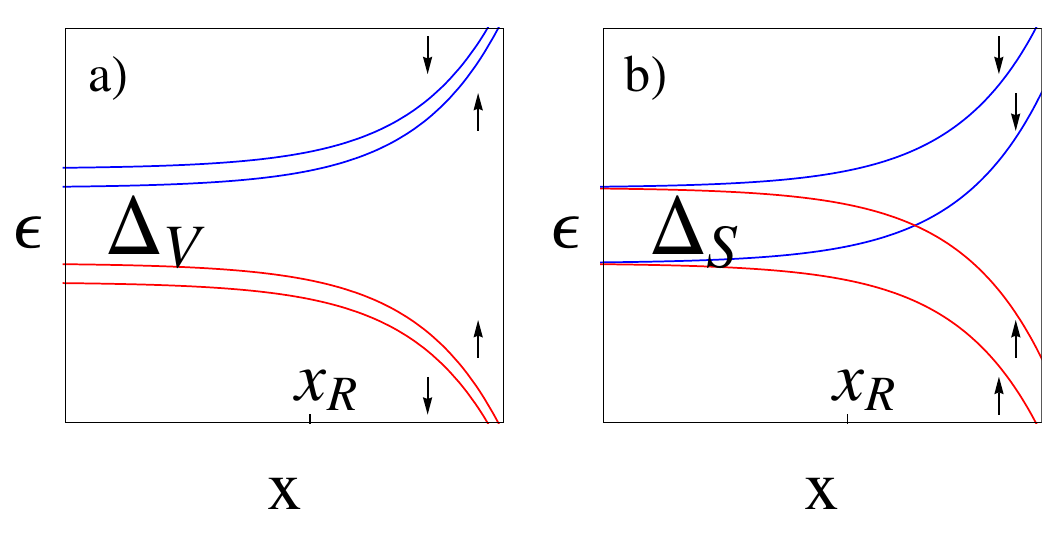}
\caption{Schematic representation of the energy dispersions of edge states of (a) valley-split ($\Delta_{V}> \Delta_{S}$) vs (b) spin-split ($\Delta_{S}> \Delta_{V}$) $\nu =0 $ QH state. In the valley-split scenario (a) there are no gapless edge states and the system resembles an insulator. Whereas the spin-split scenario supports counterpropagating spin-polarized edge states with the system identical to the quantum spin Hall state.}
\label{figsix}
\end{center}
\end{figure}
\indent
The unconventional transport phenomenology of the $\nu =0$ QH state has drawn interest from a number of theorists. Most work can be divided into two camps, one of which concentrates on the edge state structure whereas the other focuses on the bulk properties of the $\nu =0$ state. Due to the inherent particle-hole symmetry of the Dirac equation, the edge states in graphene\cite{breyfertigedge} are remarkably different from those in conventional 2DGEs. Near the physical edge of the sample, the particle-like branches disperse upwards in energy whereas the hole-like branches disperse downwards as indicated in Fig. \ref{figfive}. For the $N=0$ Landau level particle-hole symmetry dictates that each spin has one electron-like and one hole-like edge state branch. In the presence of interaction induced gaps the edge state structure would be different for a valley- or spin-split $\nu=0$ QH state. The valley-split state can be truly insulating when the Fermi energy lies in the gap since there would be no gapless edge states as shown in Fig. \ref{figsix}a. On the other hand a spin-split state would support counter-propagating edge states indicated in Fig. \ref{figsix}b, \cite{breyfertigedge,abaningeim} a scenario identical to that of quantum spin Hall systems.~\cite{HasanKane} The unusual transport phenomenology of the $\nu =0 $ QH state was initially explained assuming an interaction enhanced spin-split gap.\cite{abaningeim} However the observation of the field induced insulating feature cast some doubts on the counter-propagating edge state picture. Shimshoni {\it et. al.}\cite{shimshoni} proposed that scattering between the counterpropagating edge modes due to the presence of localized moments can lead to a metal-insulator quantum phase transition, saving the spin-polarized scenario. In this picture the crossover from metallic to insulating behavior follows a quantum KT type scaling in the magnetic field.

The observation of the a divergent resistance in Ref \onlinecite{graphene_ong} along with the KT scaling raised questions about the completeness of the edge state picture. Based on the experimental results it was argued the the $\nu =0$ QH state must exhibit XY valley-pseudospin order, where phase coherence is established between valley ${\bf K}$ and ${\bf K'}$.~\cite{valleyXY} It is well known that this state supports vortex and anti-vortex topological excitations. It was argued that in this case unbinding of the vortex-anitvortex pairs triggers a {\em bulk} KT transition from the pseudospin XY quasi-long range ordered phase to the disordered phase at a critical magnetic field $B_{c}$.

Most theoretical studies of the $\nu =0 $ QH state have been performed based on the Dirac-equation continuum models. These recent experiments raise the suspicion that the ground state for the $\nu =0$  QH state could be influenced by
physics that is outside of the Dirac-equation continuum model. Calculations based on $\pi$-band-tight-binding model including electron-electron interactions performed by one of us indicate that the ground state for graphene sheets on substrates is most likely a spin-density wave, and that a charge-density-wave is possible for suspended samples.~\cite{jeiljungpapers}
These states are favored by inter-valley exchange interactions.

It was pointed out in Ref. \onlinecite{dassarmayang} that one way to distinguish between edge and bulk mechanisms of this apparent metal insulator transition is to study samples with Corbino geometry, in which edge states play no role. Thus the same transition should be observed in Corbino samples as magnetic field increases, if the transition occurs in the bulk. On the other hand if the transition is only an edge phenomenon, then Corbino samples should only exhibit insulating behavior. Ref. \onlinecite{dassarmayang} further discusses the possible role that disorder plays in a bulk transition: disorder induces inhomogeneous charge distribution, which results in puddles with different densities and thus different filling factors in the presence of magnetic field; edge states separating different puddles may play an important role in {\em bulk} transport. Increasing magnetic field suppresses such puddle by increasing spacing between different Landau levels; this provides a possible bulk mechanism of the transition. This scenario appears consistent with experimentally observed charge-inhomogeneity in the presence of magnetic field.\cite{martin}


\subsection{Interaction induced integer QH states in bilayer graphene}
 The concept of QHF is quite general and applies whenever Landau levels with some internal flavor indices are degenerate. The
 $N\neq 0$ Landau levels of bilayer graphene have the same spin/valley four-fold degeneracy
 as single layer graphene and can be described by an approximate SU(4) invariant Hamiltonian.
The ground states corresponding to the interaction induced plateaus are then captured by the variational wavefunction (\ref{wavefunction}). Generally the physics associated with these states should be similar to that of the higher Landau levels in
single layer graphene, with the same number of collective modes and the presence of Skyrmions. However there will be some important quantitative differences related to bilayer graphene's orbital structure (see section IV A. Eq.~\ref{magfieldestate}). This is because the spatial structure of the electron-electron interactions is modified due to the $J=2$ chiral nature of bilayer graphene's charge carriers.~\cite{nomura} In particular Skyrmions are the lowest energy charge excitation for $|N| \leq 2$ in bilayer graphene. One important difference between single and bilayer graphene is that in the bilayer case it is easy to generate a mass terms (an energy difference between sublattices $A_{1}$ and $A_{2}$) externally,
simply by applying an electric field between the layers.~\cite{Yacoby}
In the $N\neq 0 $ Landau levels this should allow the study of transitions between the different broken symmetry states as a function of an external control parameter. \\
\indent
The most interesting aspects of bilayer graphene are associated with the interaction driven states
that form between $\nu = -4$ and $\nu = 4 $ QH plateaus,~\cite{bg_qhf1,bg_qhf2}
many of which do not have a counterpart in either graphene or other semiconducting QH systems.
The $N=0$ Landau level manifold in bilayer graphene has an extra pseudospin corresponding to the $n=0$ and $n=1$ orbital degree of freedom.~\cite{mccannfalko} The ordering and fluctuations of this pseudospin degree of freedom is unique to bilayer graphene and leads to unusual collective modes,~\cite{ourbilayerqhf} anomalous exciton condensation,~\cite{excitoncondbg} and spiral ordering.~\cite{coteDMbg} It is this new aspect associated with the physics of bilayer graphene that will be the focus of the rest of this section. \newline
\indent
The $N=0$ Landau level of neutral bilayer graphene is the direct product of three $S=1/2$ doublets: real spin, {\em which layer} pseudospin as in semiconducting double well quantum Hall systems (DWQHS) and the Landau level pseudospin $n=0$ and $n=1$ which is responsible for the new physics. When the magnetic field is strong enough or disorder is weak enough, interactions lead to quantum Hall plateaus at the octet's intermediate integer filling factors both for suspended bilayers~\cite{bg_qhf1} and on SiO substrates.~\cite{bg_qhf2,tutuc} For neutral bilayers the octet filling, proceeding in integer increments starting from the filling $\nu =-4$, can be characterized by Hund's rule behavior of the pseudospin: first maximize spin-polarization, then maximize layer-pseudospin to the greatest extent possible and finally maximize the Landau level pseudospin polarization to the extent allowed by the first two rules.~\cite{ourbilayerqhf} For balanced bilayers interactions favor states with spontaneous coherence between the valleys, and hence layers, since these broken symmetry states do not require charge transfer between the layers.~\cite{fertigbilayer} The Hund's rule behavior indicates that all odd integer filling factors between $\nu=-4 $ and $\nu=4$ have Landau level pseudospin polarized. This preference to polarize the $n=0$ Landau level pseudospin is because of a larger gain in exchange energy over the $n=1 $ Landau level pseudospin and is related to the wavefunction profile of the $n=0$ and $n=1$ Landau level.~\cite{ourbilayerqhf} \\
\indent
The octet state in bilayer graphene supports $\nu_{N}(8-\nu_{N})$ collective modes with $\nu_{N}$ defined similar to that in section V A. The orbital fluctuations associated with the Landau level-pseudospin are dominant at odd fillings of the octet. Exchange energy consideration at even integer fillings require that the Landau level-pseudospin degree of freedom gets frozen out, hence the interaction driven states are more similar to the semiconducting DWQHS. One particularly interesting and unusual collective mode associated with the balanced bilayer at the odd fillings is the intra-Landau level mode.~\cite{ourbilayerqhf} This collective mode exhibits a gapless $\omega_{q} \sim q^{3/2}$ dependence in the long wavelength limit, a behavior that is due to the presence of long range dipolar interactions associated with the Landau level pseudspin fluctuations. Fluctuating Landau level pseudospinors are a linear combination of $n=0 $ Landau level orbitals (even with respect to their cyclotron orbit center) and $n=1 $ orbitals (odd with respect the orbit center), and therefore carry an electric dipole proportional to the in-plane component of their pseudospin. Since the dipolar interactions are long ranged they play a dominant role at long wavelengths leading to the non-analytic long wavelength behavior.~\cite{ourbilayerqhf} The application of Kohn's theorem within the projected lowest Landau level leads to a gapless mode due to a vanishing contribution from electron-electron interactions, hence this intra-Landau level mode would acquire a gap in the presence of a single particle induced Landau level-splitting. It was proposed in Ref.~\onlinecite{ourbilayerqhf} that this mode could be observed in cyclotron resonance signal by applying an external potential difference between the layers. The external potential leads to Landau level-splitting which is positive ($n=1$ above $n=0$ orbital) in one valley and negative ($n=1$ below $n=0$ orbital) in the other.~\cite{mccannfalko}\\
\indent
The presence of the Landau level-pseudospin also has important consequences on the properties of the exciton condensates at odd fillings $\nu=-3,1$. The superfluid density vanishes at these filling factors leading to a finite-temperature fluctuation-induced first order isotropic-smectic phase transition when the layer densities are not balanced. This vanishing superfluid stiffness leads to a quadratic rather than expected linear~\cite{sds} phonon collective mode. The transition to the smectic state is a consequence of the negative Landau level-pseudospin gap in one of the valleys along with the zero superfluid stiffness; combined this leads to a long-wavelength instability of the Brazovskii-type.~\cite{brazovskii,swifthonenberg} At other odd filling factor $\nu=-1,3$, due to a negative Landau level-gap~\cite{mccannfalko}, the single particle Landau level splitting competes with the interaction induced Landau level exchange splitting which prefers $n=0$ Landau level polarization. This leads to a sequence of transitions from i) an inter-layer coherent state, ii) a mixed state with inter-orbital and inter-layer coherence to iii) an inter-orbital coherence state spiral Landau level-pseudospin ordering.~\cite{coteDMbg} The spiral state is due to the presence of Doyzaloshinskii-Moriya terms which is related to the inversion symmetry breaking of the QH ferromagnet. The presence of orbital coherence leads to a finite density of in-plane electric dipoles~\cite{shizuya} which can be manipulated by in-plane electric fields.\\
\indent
Another interesting aspect associated with the broken symmetry $N =0$ octet states in bilayer graphene is the possibility of new and interesting topological excitations not expected in graphene and semiconducting 2DEGs. It has been proposed that charge-2e Skyrmions are the low energy charged excitation near the filling factors $\nu = -2,2$.~\cite{abaninskyrmionbg} As expected these charge-2e Skyrmions have textures in both $n=0$ and $n=1$ Landau level orbitals. Interlayer charge-2e meron and Skyrmions crystals with checkerboard patterns for balanced and unbalanced layers have subsequently been identified in numerical calculations.~\cite{coteskyrmionbg} These charge-2e objects also condense in a rich variety of crystal states at odd filling factors where these pseudoskyrmions are associated with the Landau level-orbital degree of freedom. The orbital Skyrmions exhibit an unusual vorticity and charge relationship exhibiting textures of in-plane electric dipoles and can be seen as an analogue of spin Skyrmions which carry a magnetic textures. These texture would couple to an in-plane electric field and the modulation of the electronic density in the crystalline phases should be experimentally accessible through a scanning tunneling microscope measurement of their local density of states.\\
\indent
Just like monolayer graphene $\nu = 0 $ state in bilayer graphene has drawn special attention. In the quantum Hall regime the ordering seems to indicate either spin or valley polarization, or both.~\cite{bg_qhf1,bg_qhf2} The interest in the $\nu =0$ QH state of bilayer graphene is different from that in graphene though, as the parabolic dispersion of bilayer graphene makes it susceptible to interaction-induced symmetry breaking even at zero magnetic field.~\cite{Min,gapped2,gapless} Incompressibilities for suspended bilayer graphene were seen to survive even in the $B \to 0$ limit,~\cite{Yacoby,Lau} where it was concluded that the incompressibility at the charge neutrality was consistent with either spontaneously broken time-reversal symmetry  or spontaneously broken rotational symmetry. The broken time reversal symmetry leads to an anomalous quantum Hall state where electrons in one layer occupy valley ${\bf K}$ whereas electrons in the other layer occupy valley ${\bf K'}$.~\cite{gapped2} This state supports topologically protected edge modes and would exhibit a finite conductance. The time-reversal symmetric state is predicted to exhibit nematic order where the parabolic dispersion is expected to split into two Dirac cones. This would result in a lowering of the density of states at the charge neutrality point.~\cite{gapless} \\
\indent
Some authors\cite{Min,Levitov} have concluded that the $B=0$ broken symmetry states
have spin and valley dependent sublattice polarization.
If this is true the states are associated with interesting momentum space Berry
curvature,\cite{NewNiu} anomalous Hall, and orbital magnetism effects.
Among the states in this class with no overall layer polarization,
are\cite{NewZhang} SDW states, usually called LAF states (layer antiferromagnet) in the bilayer case,
QSH (quantum spin Hall) and QAH\cite{Levitov} (quantum anomalous Hall states).  These three states
have quantized Hall conductances with the values $0$, $0$, and $\pm 4 e^2/h$ for the
LAF, QSH, and QAH states respectively.  In each case the gaps associated with order evolve
smoothly into quantum Hall gaps with filling factors $\nu$ = $0,0,4$ respectively.
In a magnetic field the QAH state evolves into a bilayer state that is similar to the non-interacting
$\nu=4$ state.  The QSH state evolves in field into a spin-polarized state with total filling factor $\nu=0$ but
a filling factor difference of 4 between majority and minority spins.  The LAF state evolves into a state which
appears at $\nu=0$ and has no spin-polarization.  Given these properties, studies of
the quantum Hall effect at weak magnetic fields can shed light on the character of the $B=0$
broken symmetry states.  For example if the QAH state is the $B=0$ ground state\cite{Levitov} the
$\nu=4$ quantum Hall effect should persist to $B=0$, whereas if the LAF state where the ground state
it is the $\nu=0$ quantum Hall state that would persist. Current experiments do not yet provide consistent results on these points, presumably
because of small differences in disorder potential.  The possibility of observing quite
different transport properties on quite similar samples suggest that the various potential
ordered states of bilayers compete closely.\\

\section{Fractional Hall Effect in Graphene}
As explained earlier electron-electron interactions in semiconducting 2DEGs can give rise to incompressibilities at fractional values of the Landau level filling factor $\nu$. Interactions within a partially filled Landau level lead to the formation of strongly correlated states that exhibit fractionally charged excitations and fractional statistics~\cite{fqheth} which in some cases are proposed to be non-Abelian.~\cite{nonabelianrmp} Graphene is very similar to its semiconducting 2DEG cousin; in fact there are more similarities than differences as far as FQHE are concerned. The two dimensionality, Landau level structure and the appearance of integer QH effects due to both Landau quantization and interaction induced gaps are good benchmarks indicating that incompressibilities at fractional fillings are likely to exist.  While similarities do exist there are also differences, due to the added valley degeneracy and the possibility of tuning Coulomb interactions strength by substrate manipulation. As a result of these features FQHE in graphene is expected to be richer. Specifically the four-fold degeneracy in graphene can lead to spontaneously broken $SU(4)$ symmetric FQH states with an ordering of internal indices that lead to a multitude of FQH states not observed or anticipated in semiconducting 2DEGs.\cite{ydm}\\
\indent
FQH physics is expected to dominate when electron-electron interactions are stronger than disorder potentials. At the time of this writing it still appears to be true that this ratio has a larger value in some semiconducting quantum well systems than in the cleanest available graphene samples. Current theories of FQHE in semiconducting systems rely heavily on the notion of the projection of interactions onto a single Landau level (i.e in general cases Landau level mixing is neglected). The appropriateness of the Landau level projection is not trivially obvious for the case of graphene. In semiconducting 2DEGs this is formally achieved in the limit of a large magnetic field $B \to \infty$. This is justified because interaction gaps $V_{e-e} \sim e^2/\varepsilon l_{B}$ scale as $\sqrt{B}$ while the single particle Landau level gap $\hbar \omega_{c}$ scales as $B$. So it can be argued there that for large magnetic fields interactions between electrons in the lowest partially-filled Landau level cannot induce transitions to the higher Landau levels. The projection of the interactions onto the lowest partially-filled Landau level is not as clearly justified in graphene because the single particle band gaps in graphene also scale as $ \sim \sqrt{B}$, which is the same as the interaction strength. However, the validity of Landau level projection in theories of the graphene FQHE can be justified by considering the large dielectric constant limit in the surrounding material. Most FQHE theory for graphene has nevertheless assumed that Landau level mixing is an inessential complication that can be ignored.\\
\indent
On the experimental side, the focus in graphene has shifted from IQHE to FQHE since higher mobilities have been achieved both by suspending samples and using h-BN as a substrate. The FQHE was initially observed in suspended samples via two terminal transport measurements which revealed clear plateaus corresponding to the $\nu = 2-5/3 =1/3$ FQH state developing as a function of the magnetic field.~\cite{graphene_fqhe1,graphene_fqhe2} More recently a sequence of FQH states have been observed in high mobility multi-terminal graphene devices fabricated on a h-BN substrate.~\cite{BNsubstrate} The observed fractions are all thirds (i.e. of the form $\nu = k/3$) and belong to the $N=0$ and $N=1$ graphene Landau levels. In the $N=0$ Landau level, the observed fractions indicate multicomponent FQH states with spontaneously broken $SU(4)$ symmetry, and possibly full symmetry broken states in $N=1$ Landau level.
It is interesting to note that none of the fractions currently observed in the $N=0$ Landau level of graphene have an analogue in the semiconducting GaAs; for example none of the observed fractions correspond to the celebrated Laughlin state at $\nu_N =1/3$. The observed fractions are most likely multicomponent generalization of the Laughlin state. Variational wavefunctions with an internal $SU(2)$ symmetry were introduced by Halperin by generalization of the Laughlin state to spin-1/2 particles. In the following we review the essential properties of the Laughlin state and provide the $SU(4)$ generalization of these states which are most likely related to the FQH states observed recently.  \\
\indent
The FQHE is characterized by the condensation of electrons into special highly correlated states that minimize the Coulomb energy. Interactions play a crucial role in determining the nature of these correlated states. A fractionally filled Landau level of noninteracting electrons does not define a unique many-body state since electrons can be placed into any of the degenerate orbitals in an exponentially large number of ways subject only to the Pauli principle. Interactions lift this degeneracy by selecting a unique ground state from the vast space of degenerate states, in which electrons avoid each other as much as possible. At some specific filling factor $\nu_N$, the energy change upon adding one electron differs from the energy change for removing one electron by a finite charge gap. The resulting fluid is incompressible and often characterized by unusual topological order.~\cite{addrefgirvinahm} For fractional fillings of the form $\nu_N=1/m$ the essence of this phenomenon is captured by Laughlin's wavefunction,
\begin{equation}
\psi_{m}[z_{i}] = \prod_{i<j} (z_{i} - z_{j})^{m} \exp(-\frac{1}{4}\sum_{i} |z_{i}|^2),
\end{equation}
where $z_{i} = x_{i} + i y_{i}$ denotes the two-dimensional positions of the electron in terms of a complex number, and we have set the magnetic length $l_{B} =1$. Since this wavefunction lives in the lowest Landau level (for $N=0$, corresponding wave functions exist for any $N$) the parameter $m$ must be an integer to satisfy wavefunction analyticity.~\cite{girvinjach} Furthermore to satisfy the wavefunction antisymmetry requirement for fermions, $m$ must be odd. The Laughlin wavefunction has good correlations built in it since each electrons sees $m$-zeros at the positions of all the other electrons. This helps minimize the Coulomb interactions uniformly among all the electrons. The probability of two electrons approaching each other, which is given by the amplitude of the wavefunction, vanishes rapidly: $ p({\bf r}_{i} - {\bf r}_{j})\sim |{\bf r}_{i} - {\bf r}_{j}|^{2m}$. Note that no  pair of particles ever has relative angular momentum less than $m$, and therefore the Laughlin state is a zero-energy exact eigenstates of model Hamiltonians initially introduced by Haldane in which electron only interact if they have small relative angular momentum. In the following we briefly review this point of view on FQH physics for semiconducting 2DGEGs and apply these ideas to describe wavefunctions for $SU(4)$ FQHE in graphene.\\
\indent
In semiconducting 2DEGs the primary features of FQH states are captured by
Haldane pseudopotentials $V^{(n)}_{m}$ which represent the interaction energy of two electrons with relative angular momentum $m$. Haldane's pseudopotentials are essential for exact diagonalization calculation for small sized systems and can give insight into possible FQH states by looking at overlaps of trial wavefunctions with exact numerical solutions. Furthermore this procedure introduces model Hamiltonians for which exact eigenstates which exhibit incompressibility at fractional filling are available. An example of this is the hard core model for which the zero energy eigenstate is the celebrated Laughlin state. Any given interaction $V(|{\bf r}_{i}-{\bf r}_{j}|)$ can be projected onto relative angular momentum basis $| m' \rangle$ where $m'$,
\begin{equation}
H = \sum_{i<j} \sum_{m'} V_{m'} P^{m'}_{ij},
\end{equation}
where
\begin{equation}
V_{m'} = \frac{\langle m'|V|m' \rangle}{\langle m'|m'\rangle}.
\end{equation}
are known as the Haldane pseudopotentials and $P^{m'}_{ij}$ is the projection operator which selects particles with relative angular momentum $m'$. The Laughlin state is the exact zero energy eigenstate of the "hard-core" potential with $V_{m'} = 0 $ for $m' \geq m$. This simply follows from the fact that $P^{m'}_{ij} |\psi_{m}  \rangle  =0 $ for any $m' < m$ since every pair has relative angular momentum of at least $m$. These models allow for a simple generalization to the case of multicomponent wavefunctions. \\
\indent
By mapping Hamiltoninans projected onto different isolated Landau levels, Haldane pseudpotenetials can be calculated for any value of $N$. In the $N=0$ Landau level in graphene the wavefunctions are localized on sublattice $A (B)$ for valley ${\bf K}({\bf K'})$ with orbital index $n=0$, whereas for the $N \neq 0$ graphene Landau level the wavefunction is a coherent superposition of $n$ and $n-1$ orbital wavefunction of semiconducting 2DEGs localized of different sublattices. It follows that~\cite{nomura,gmd} the $N=0$ Landau level graphene pseudopotentials are identical to the lowest Landau level pseudopotentials in semiconducting 2DEGs, however the $N \neq 0$ pseudopotentials are different and given by,
\begin{equation}
V^{(N)}_{m} = \int \frac{d^{2}q}{(2 \pi)^{2}} v_{q} e^{-|q|^{2}}
L_{m}(q^{2})\bigg[ \frac{1}{2}[L_{|N|}(q^{2}/2) + L_{|N|-1}(q^{2}/2)] \bigg]^{2},
\end{equation}
where $L_{n}(x)$ is the $n^{th}$ order Laguerre polynomial and $v_{q}$ is the Fourier transform of the interaction potential with $v_{q}= 2\pi e^2/(\epsilon q)$ for Coulomb interaction. The relative strength of the Haldane pseudopotential $V^{N}_{1}/V^{N}_{3}$ normally indicates the possibility of finding strongly correlated FQH states. Numerical investigations have shown that graphene should support FQH-like correlated states in the lowest Landau level.~\cite{zlatkogoerbig} \newline
\indent
Motivated by Haldane pseudoptential considerations one can write a class of model Hamiltonians with exact zero-energy eigenstates.
Relevant to graphene are the $SU(4)$ generalization of the Halperin-Laughlin wavefunctions~\cite{goerbigFQHE}
\begin{equation}
\label{HLwave}
\Psi^{SU(4)} = \prod_{\alpha=1}^{4} \prod_{i_{\alpha}< j_{\alpha} } (z^{\alpha}_{i_{\alpha}} -z^{\alpha}_{j_{\alpha}})^{m_{\alpha}} \prod_{\alpha < \beta}^{4} \prod_{i_{\alpha},j_{\beta}}
 (z^{\alpha}_{i_{\alpha}} -z^{\beta}_{j_{\beta}})^{n_{\alpha \beta}},
\end{equation}
where we have omitted the ubiquitous Gaussian factor. In the above equation $z^{\alpha}_{j_{\alpha}} = x^{\alpha}_{j_{\alpha}} + iy^{\alpha}_{j_{\alpha}} $ denotes the position of the $i^{th}$ particle with a flavor index $\alpha $. It is easy to write the relevant model Hamiltonians for which these wavefunctions are the zero-energy eigenstates, we leave this as an exercise for the reader. Apart from the omitted exponential factor, the wavefunction must be a homogeneous polynomial of degree $N^{\alpha}_{\phi}$ in $z^{\alpha}_{i_{\alpha}}$ where $N^{\alpha}_{\phi}$ is the degeneracy of of the $\alpha$ Landau level. Furthermore the polynomial must be antisymmetric under exchange of  $z_{i_{\alpha}}$ and $z_{j_{\alpha}}$ for the same index $\alpha$, which restrict $m_{\alpha}$'s to take odd integer values. Using the fact that every extended single particle orbital in the lowest Landau level involves a polynomial with a fixed density of zeros given by $B/\Phi_{0}$ (where $\Phi_{0} \equiv hc/e$ is the flux quantum) one can express the density $\rho_{\alpha}$ of each $\alpha $ component $B/\Phi_{0} = \rho_{\alpha} m_{\alpha} + \sum_{\alpha \neq \beta}  \rho_{\alpha} n_{\alpha \beta}$. The individual filling factors can be calculated using
\begin{equation}
\label{filling}
(\nu_{1},\nu_{2},\nu_3,\nu_4 )^{T} = K^{-1}(1,1,1,1)^{T} ,
\end{equation}
by defining a symmetric $K$ matrix with the diagonal elements $k_{\alpha \alpha} = m_{\alpha} $ and the off-diagonal components $k_{\alpha \beta} = n_{\alpha \beta}$. The total filling factor corresponding to $K$ is given by $ \nu_{N} = \sum_{i} \nu_{i}$. Eq(\ref{filling}) only makes sense if the matrix K is invertible. One can classify the spin and valley polarization of these states based on the rank of the corresponding K matrix and the filling factors. For example since a rank 4 matrix is invertible the individual spin and valley flavor filling factors $ \nu_{i}$ are uniquely determined, therefore the spin polarization in the z-direction is just given as a $S^{z} = N/2 \sum_{\sigma} (\nu_{\sigma,\uparrow} - \nu_{\sigma,\downarrow})$ with a similar expression for the valley polarization.~\cite{goerbigreview} The single component $\nu =1/m$ Laughlin like states correspond to $K$ matrices of rank 1 with all the exponents $m_{\alpha} = n_{\alpha,\beta} = m$, such a state would be a fully polarized SU(4)-ferromagnetic state. The consideration of spin and valley ordering is important in determining the ground state for a filling factor as quantum Hall ferromagnetic physics prefers states that exhibit ferromagnetic ordering. The $K$-matrix can be viewed as a SU(4) generalization of Wen's K-matrix formulation of the FQH effect. This $K$-matrix also encodes the quasiparticle statistics and charge for a given state.~\cite{WenZee}\\
\indent
From the experimental results on h-BN substrate it seems that these SU(4) states describe the observed fractional plateaus within the lowest Landau level in graphene. It seems that in the lowest Landau level the explicit symmetry breaking terms including valley and Zeeman splitting are sufficiently small that the Coulomb interactions intermix the four branches, allowing for mixed valley-spin FQH states.~\cite{BNsubstrate} One can argue this based on the pattern of the particle-hole symmetry of the observed FQH states; for example particle-hole conjugate FQH state must share similar gaps. The observation of the fractional states $\nu = 1/3,2/3$ and $4/3$ and their conjugate fractions seem to indicate that the external symmetry breaking terms are sufficiently small and an approximate SU(4) symmetry is preserved. The FQH states observed in the $N=1$ Landau level do not follow this patterns indicating that all degeneracies are lifted either due to explicit symmetry breaking terms or coupling with external fields. Initial investigation seem to reveal that the fractional states in the $N=1$ Landau level are direct analogues of the Laughlin state observed in GaAs. The $1/3$ states in second Landau level are fairly robust unlike the case of GaAs, this is to be expected as the relativistic form factors for $N=1$ lead to a more stable state when compared with the lowest Landau level $1/3$-state in graphene.~\cite{chakraborty} \\
\indent
As the sample mobilities increase it is expected that additional plateaus associated with the hierarchical states will also appear. A more interesting aspect would be the appearance of some non-Abelian FQH states such as the Pffafian or Read-Rezayi states.~\cite{nonabelianrmp} Unlike semiconducting 2DEGs, modification of interaction strength in graphene due to substrate manipulation could allow for tunability and hence stability of several FQH states. It has been proposed that modification of the dielectric environment could lead to stability of the Pffafian and Laughlin states.~\cite{zlatkodima} The chiral band structure of massive graphene and bilayer graphene also allows for the possibility of tunable quantum phase transitions between incompressible and compressible states along with subtle transitions between incompressible states characterized by different topological orders.~\cite{massiveDFQHE} Furthermore since a metallic gate is used to vary the doping concentration an image charge accompanies the electron leading to electric dipole-dipole interactions between the carriers which can be tuned by varying the distance of the the gate from the graphene layer. This dipolar interaction will also have important consequences on the edge states in particular with regards to the edge reconstruction associated with FQH states of graphene.~\cite{kuninprep} The important issue of FQH edge reconstruction has been previously discussed both in the semiconductor 2DEG\cite{edgereconstruction} and graphene\cite{casronetoedgereconstruction} contexts.\\
\indent

\section{Conclusions and Outlook}

By demonstrating that a Landau level is pinned to the carrier neutrality point,
quantum Hall studies of graphene's two dimensional electron system have in
a very direct way confirmed the massless Dirac character of its
bands.  Similar studies have already also confirmed the massive Dirac character of
bilayer graphene bands and are making progress toward revealing the distinct bands of
ABA and ABC trilayers.~\cite{trilayerIQHE}  Indeed the few-layer family of graphene 2DEGs has
impressive variety which can be accessed by varying the number of layers and how they are stacked.
Electronic properties at low energies can depend on inter-layer coupling details that
are not necessarily predictable with sufficient accuracy using {\em ab initio} electronic
structure theory.  We expect that sample characterization based on Shubnikov-de Haas
oscillations and quantum Hall effects will continue to prove valuable in
accurately characterizing the low-energy band structures.  Although this practical aspect of
quantum Hall measurements is still important, studies of
regimes in which interactions play an essential role are potentially
even more valuable because they may reveal interesting new physics.

Quantum Hall studies have been shedding light on
interaction physics of these systems in three different ways.
First of all, the pattern of
quantum Hall effects as the weak magnetic field limit is approached sheds
important light on the zero-field ground state.  This is particularly true of $B=0$ states
that have sublattice polarization and associated momentum space Berry curvatures that
lead to quantized anomalous Hall effects with Hall  conductivities  (given in $e^2/h$ units)
$0$, $\nu=\pm 1$, and $\nu=\pm 2$ in
single layer graphene, and $0$, $\pm 2$, and $\pm 4$ in bilayer graphene.
When any of these states occur, the gap is expected to persist to finite fields and to occur at Landau
level filling factor equal to the total Hall conductance integer which characterizes the $B=0$
state.  As the field strengthens the transport phenomenology of the gapped state is expected to cross over to that of an ordinary Hall plateau.  
If the ground state of single layer graphene was a CDW state,
a $\nu=0$ quantum Hall effect would be expected at all field strengths.  The experimental observation that
the $\nu=0$ graphene quantum Hall effect does not survive to $B \to 0$, seems to establish
unequivocally that the $B=0$ ground state is not a CDW.  In bilayer graphene, on the other hand,
many experimental hints are emerging that quantum Hall effects may survive to $B=0$,
although this question is just now in the midst of being clarified experimentally.
Persistent $\nu=\pm 4$ quantum Hall effects at low fields suggest stable quantized
anomalous Hall states whereas persistent $\nu=0$ states at low fields suggest
antiferromagnetic states with opposite spin-polarizations on opposite layers.
The quantum Hall physics of graphene trilayers, which we have not reviewed extensively here,
is still at an earlier stage of exploration.  It is nevertheless clear that ordered electronic states are likely even at
$B=0$, especially in ABC trilayers, and that anomalously persistent weak-field quantum
Hall effects will if present play a valuable role in sorting out their character.  \\
\indent
Interactions in C2DEGs also play a prominent role by generating additional integer quantum Hall plateaus
that are not expected by Landau quantization alone but are unrelated to possible $B=0$
spontaneous Hall states.  Instead they can be understood by considering interactions among electrons
that share the same well-resolved Landau level.
Recent experiments on graphene samples on h-BN substrates seem to suggest that interactions within the four-fold degenerate Landau levels lead to the formation of interaction induced gaps and hence additional quantum Hall effects. If these states spontaneously break the SU(4) symmetry then the possibility of observing exotic topological excitation associated with SU(4) Skyrmions remains intriguing. Recent experimental studies indicate that the $\nu =0 $ QH plateau in graphene is an insulating state, indicating that it is most likely some sort of a density wave ordered state.   More unusual states are likely
to be associated with Hall plateaus that appear within the $N=0$ Landau levels of C2DEGs with a chirality index $J \geq 2$. In these cases the zero-energy Landau level supports multiple orbital degrees of freedom along with the four-fold degeneracy already present due to spin/valley.   Speaking loosely, quantum states corresponding to cyclotron orbits with different radius,
which would have different energies in an ordinary two-dimensional electron gas, are degenerate.
Experiments have confirmed the role of interactions in bilayer graphene as additional integer quantum Hall plateaus due to interaction induced gaps within the eight-fold degenerate zero-energy Landau level have been observed by several groups. When the Landau level orbital degree of freedom is active (for example at odd integer filling factors in $J=2$ bilayers)
Landau level orbital ordering reveals exotic QH states, which in the pseudospin language are analogs of spiral and helical ferromagnets.
Because the quantum Hall effects at odd integer filling factors are least strong and more
easily altered by disorder, they have not yet been studied extensively.
Transport experiments in separately contacted bilayer graphene, and
optical experiments are likely to yield many surprises in the future.
Initial theoretical investigations in ABC-stacked trilayer graphene, which at the relevant energy scale belongs to the class of $J=3$ C2DEGs, also indicate that integer induced quantum Hall plateaus which support one-dimensional charge density waves with broken translational symmetry appear at certain integer filling factors.~\cite{YBtrilayers}
These type of charge density wave states at integer filling factors have no direct analogue in semiconducting 2DEGs and are a consequence of interactions between electrons in different Landau level orbitals.  The study of quantum Hall ferromagnets in
bilayers and trilayers is still in its infancy and appears capable of revealing new physics that has not yet
been anticipated.   \\
\indent
Finally of course, interactions in graphene systems of sufficiently high quality are expected to give rise to
quantum Hall effects at a large number of fractional filling factors.
Recent improvements in graphene sample mobility have enabled the
first observations of FQH states in graphene.
The FQH states observed so far in the $N=0$ Landau level are most likely multicomponent generalizations
of the Laughlin states that are prominent in ordinary two-dimensional electron gases.
The competition between incompressible states and disorder is certainly influenced by the
difference in Landau level degeneracy, and this could be important in explaining the
fractions which have been observed to date.
As expected, due to the orbital character of the graphene which influences the Haldane pseudopotentials, the FQH states in the $N=1$ Landau level are more robust than the FQH states in $N=0$ Landau level.~\cite{chakraborty,massiveDFQHE}
The $N=1$ FQHE in graphene is different from $N=0$, and also from $n=1$ in an ordinary 2DEG, because of differences in the interaction form factors.
An especially exciting possibility is the observation of non-Abelian FQH states in graphene,
which could end up being more robust than they are in the ordinary 2DEG case due to the advantage of tunable interactions in C2DEGs,
as suggested by recent theoretical studies.~\cite{zlatkodima,massiveDFQHE}
Both the filling factors at which incompressible states appear, and the character of their wavefunctions
can change as a direct result of the chiral nature of charge carriers in graphene and bilayer graphene.
Searches for correlated FQH states in the $N=0$ Landau level of bilayer graphene
with its additional orbital degeneracy is quite possibly the most exciting avenue and very likely to yield many surprises.
Progress to date has been hampered by the numerical challenges posed by the additional degeneracies, and
by the lack of predictive theoretical insights.

This work was supported by the State of Florida (YB), by NSF grant DMR-1004545 (KY), and
by Welch Foundation grant TBF1473 (AHM).

\end{document}